\documentclass{article}
\usepackage[utf8]{inputenc}

\usepackage{gensymb}
\usepackage{amsfonts}
\usepackage{amsmath}
\usepackage{mathtools}
\usepackage{cite}
\usepackage{easybmat}

\usepackage{amsthm}

\newtheorem{design}{Design Criteria}
\newtheorem{coro}{Corollary}
\usepackage{hyperref}
\usepackage{hyperref}
\usepackage{gensymb}
\usepackage{amsfonts}
\usepackage{amsmath}
\usepackage{mathtools}
\usepackage{easybmat}
\usepackage{xcolor}
\usepackage{soul}
\newcommand{\catname}[1]{{\normalfont\textbf{#1}}}
\newcommand{\FinStat}{\catname{FinStat}}
\usepackage{amsthm}
\newtheorem{theorem2}{Theorem}
\usepackage{hyperref}
\title{The Design of Entropic Quantifiers of Correlations and Their Application To Sufficiency}
\author{Nicholas Carrara\thanks{ncarrara@albany.edu}\\ Dept. of Physics\\University at Albany\\Albany, NY 12222 \and Kevin Vanslette\thanks{kvanslet@mit.edu}\\ Dept. of Mechanical Engineering\\Massachusettes Institute of Technology\\Cambridge, MA 02139}

\date{December 2019}

\begin{document}

\maketitle

% Abstract (Do not insert blank lines, i.e. \\) 
\abstract{Using first principles from inference, we design the total correlation (TC) and $n$-partite information (NPI) functionals through a process of eliminative induction.  These quantities are designed for the purposes of \textit{ranking} joint probability distributions with respect to their correlations.  The mutual($n$-partite) information is found as a special case of the total correlation, when the variable space is split into two($n$) distinct subspaces of interest.  The design of these functionals is imposed according to a \textit{Principle of Constant Correlations} (PCC), which constrains correlation functionals to behave in a specified way when independence conditions are present.  Together with our design goal, the PCC guides us in choosing the appropriate design criteria for constructing the desired functionals.  We find, through the PCC, that the total correlation TC and $n$-partite information NPI are the unique functionals capable of determining whether a certain class of inferential transformations, $\rho\xrightarrow{*}\rho'$, preserve, destroy or create correlations.  This provides conceptual clarity by ruling out other possible global correlation quantifiers.}

\section{Introduction}\label{intro}
The goal of this paper is to quantify the notion of \textit{global correlations} as it pertains to inductive inference.  This is achieved by designing a set of functionals from first principles to rank entire probability distributions $\rho$ according to their correlations.  Because correlations are relationships defined \emph{between} different subspaces of propositions (variables), the ranking of any distribution $\rho$, and hence the type of correlation functional one arrives at, depends on the particular choice of ``split" or partitioning of the variable space.  Each choice of ``split'' produces a unique functional for quantifying global correlations, which we call the $n$-partite information (NPI).

The term correlation may be defined colloquially as being a relation between two or more ``things".  
While we have a sense of what correlations are, how do we quantify this notion more precisely?  If correlations have to do with ``things'' in the real world, are correlations themselves ``real?''  Can correlations be ``physical?''  One is forced to address similar questions in the context of designing the relative entropy as a tool for updating probability distributions in the presence of new information (e.g. ``What is information?'') \cite{Caticha_towards}.  In the context of inference, correlations are broadly defined as being statistical relationships between propositions.  In this paper we adopt the view that whatever correlations may be, their effect is to influence our beliefs about the natural world.  Thus, they are interpreted as the information which constitutes statistical dependency.  With this identification, the natural setting for the discussion becomes inductive inference.   

%\textcolor{red}{Must correlations be entirely epistemic?? Can no correlation be ``physical"? Can we not speak of ``real" correlations? If they are neither real or physical, how can we verify them? We can have epistemic uncertainty regarding the reality of correlations as we can have uncertainty in the validity of a model, but nothing prevents the existence of unknown yet "ontic" underlying correlations.}

%When one has complete information, the rules for inference are dictated by \textit{deductive} logic and the truth value of some proposition given another, can be found by constructing truth tables.  Correlation is represented by the implication arrow ``$x \Rightarrow y$.''  These relationships most often address the question: given a proposition $x$ is ``true'', what should one believe about the ``truth'' of some other proposition $y$?  In other words, ``how does information about $y$ influence our beliefs about $x$? We do this by designing a set of functionals over the space of probability distributions according to a set of design criteria (DC) which are inspired by a guiding principle.  The DC are chosen so as to restrict the behavior of the functionals in certain special cases, which leads to unique solutions of their general functional form.''   

When one has incomplete information, the tools one must use for reasoning objectively are probabilities \cite{Cox,Caticha_towards}. The relationships between different propositions $x$ and $y$ are quantified by a joint probability density, $p(x,y)=p(x|y)p(y)=p(x)p(y|x)$, where the conditional distribution $p(y|x)$ quantifies what one should believe about $y$ given information about $x$, and vice-versa for $p(x|y)$.  Intuitively, correlations should have something to do with these conditional dependencies.

In this paper, we seek to quantify a \textit{global amount} of correlation for an entire probability distribution.  That is, we desire a scalar functional $I[\rho]$ for the purpose of ranking distributions $\rho$ according to their correlations.  Such functionals are not unique since many examples, e.g. covariance, correlation coefficient \cite{Pearson}, distance correlation \cite{Szekely}, mutual information \cite{CoverThomas_Book}, total correlation \cite{Watanabe}, maximal-information coefficient \cite{Reshef}, etc., \textit{measure} correlations in different ways.  What we desire is a principled approach to designing a family of measures $I[\rho]$ according to specific design criteria \cite{ShoreJohnson,Skilling,Caticha_Book}.

The idea of designing a functional for \textit{ranking} probability distributions was first discussed in Skilling \cite{Skilling}.  In his paper, Skilling designs the relative entropy as a tool for ranking posterior distributions, $\rho$, with respect to a prior, $\varphi$, in the presence of new information that comes in the form of constraints (\ref{constraints}) (see \hyperref[entropicupdating]{Section 2.1.2} for details).  The ability of the relative entropy to provide a \textit{ranking} of posterior distributions allows one to choose the posterior that is \textit{closest} to the prior while still incorporating the new information that is provided by the constraints.  Thus, one can choose to update the prior in the most minimalist way possible.  This feature is part of the overall \textit{objectivity} that is incorporated into the design of relative entropy and in later versions is stated as the guiding principle \cite{Caticha_Entropy_1,Vanslette,VansletteThesis}.

Like relative entropy, we desire a method for \textit{ranking} joint distributions with respect to their correlations.  Whatever the value of our desired quantifier $I[\rho]$ gives for a particular distribution $\rho$, we expect that if we change $\rho$ through some generic transformation $(*)$, $\rho \xrightarrow{*} \rho' = \rho + \delta \rho$, that our quantifier also changes $I[\rho] \rightarrow I[\rho'] = I[\rho] + \delta I$, and that this change of $I[\rho]$ reflects the change in the correlations, i.e. if $\rho$ changes in a way that increases the correlations, then $I[\rho]$ should also increase.  Thus, our quantifier should be an increasing functional of the correlations, i.e. it should provide a \textit{ranking} of $\rho$'s.

The \textit{type} of correlation functional $I[\rho]$ one arrives at depends on a choice of the \textit{splits} within the proposition space $\mathbf{X}$, and thus the functional we seek is $I[\rho]\rightarrow I[\rho,\mathbf{X}]$.  For example, if one has a proposition space $\mathbf{X} = \mathbf{X}_1\times\dots\times\mathbf{X}_N$, consisting of $N$ variables, then one must specify \textit{which} correlations the functional $I[\rho,\mathbf{X}]$ should quantify.  Do we wish to quantify how the variable $\mathbf{X}_1$ is correlated with the other $N-1$ variables?  Or do we want to study the correlations \textit{between} all of the variables?  In our design derivation, each of these questions represent the extremal cases of the family of quantifiers $I[\rho,\mathbf{X}]$, the former being a \textit{bi-partite} correlation (or mutual information) functional and the latter being a \textit{total} correlation functional.

In the main design derivation we will focus on the the case of \textit{total correlation} which is designed to quantify the correlations between every variable subspace $\mathbf{X}_i$ in a set of variables $\mathbf{X} = \mathbf{X}_1\times\dots\times\mathbf{X}_N$.  We suggest a set of design criteria (DC) for the purpose of designing such a tool. These DC are guided by the Principle of Constant Correlations (PCC), which states that ``the amount of correlations in $\rho$ should not change unless required by the transformation, $(\rho,\mathbf{X})\stackrel{*}{\rightarrow}(\rho',\mathbf{X}')$.'' This implies our design derivation requires us to study equivalence classes of $[\rho]$ within statistical manifolds $\mathbf{\Delta}$ under the various transformations of distributions $\rho$ that are typically performed in inference tasks. We will find, according to our design criteria, that the global quantifier of correlations we desire in this special case is equivalent to the total correlation \cite{Watanabe}.
%\textcolor{red}{or should it be the NPI, which can take the TC and MI as special cases}

Once one arrives at the TC as the solution to the design problem in this article, one can then derive special cases such as the \textit{mutual information} \cite{CoverThomas_Book} or, as we will call them, any \textit{$n$-partite information} (NPI), which measures the correlations shared between generic $n$-partitions of the proposition space.  The NPI and the mutual information (or bi-partite information) can be derived using the same principles as the TC except with one modification, as we will discuss in \hyperref[npartite]{Section V}.

The special case of NPI when $n=2$ is the \textit{bipartite} (or mutual) information, which quantifies the \textit{amount} of correlations present between two subsets of some proposition space $\mathbf{X}$.  Mutual information (MI) as a measure of correlation has a long history, beginning with Shannon's seminal work on communication theory \cite{Shannon_Book} in which he first defines it.  While Shannon provided arguments for the functional form of his entropy \cite{Shannon_Book}, he did not provide a derivation of (MI). Despite this, there has still been no principled approach to the design of MI or for the total correlation TC.  Recently however, there has been an interest in characterizing entropy through a category theoretic approach (see the works of Baez et. al. \cite{Baez}).  The approach by Baez et. al. shows that a particular class of functors from the category \FinStat, which is a finite set equipped with a probability distribution, are scalar multiples of the entropy \cite{Baez}.  The papers by Baudot et. al. \cite{Baudot12,Baudot22,Baudot32} also take a category theoretical approach however their results are more focused on the topological properties of information theoretic quantities.  Both Baez et. al. and Baudot et. al. discuss various information theoretic measures such as the relative entropy, mutual information, total correlation, and others.  

The idea of designing a tool for the purpose of inference and information theory is not new. Beginning in \cite{Cox}, Cox showed that probabilities are the functions that are designed to quantify ``reasonable expectation'' \cite{Cox2}, of which Jaynes \cite{Jaynes_Book} and Caticha \cite{Caticha_Book} have since improved upon as ``degrees of rational belief''.  Inspired by the method of maximum entropy \cite{Jaynes_Book,Jaynes1,Jaynes2}, there have been many improvements on the derivation of entropy as a tool designed for the purpose of updating probability distributions in the decades since Shannon \cite{Shannon_Book}. Most notably they are by Shore and Johnson \cite{ShoreJohnson}, Skilling \cite{Skilling}, Caticha \cite{Caticha_Entropy_1}, and Vanslette \cite{Vanslette,VansletteThesis}. The entropy functionals in \cite{Caticha_Entropy_1,Vanslette,VansletteThesis} are designed to follow the Principle of Minimal Updating (PMU), which states, for the purpose of enforcing objectivity, that ``a probability distribution should only be updated to the extent required by the new information.'' In these articles, information is defined operationally $(*)$ as that which induces the updating of the probability distributions, $\varphi\stackrel{*}{\rightarrow}\rho$.

An important consequence of deriving the various NPI as tools for ranking is their immediate application to the notion of \textit{statistical sufficiency}.  Sufficiency is a concept that dates back to Fisher, and some would argue Laplace \cite{Stigler}, both of whom were interested in finding statistics that \textit{contained} all relevant information about a sample.  Such statistics are called \textit{sufficient}, however this notion is only a binary label, so it does not quantify an \textit{amount} of sufficiency.  Using the result of our design derivation, we can propose a new definition of sufficiency in terms of a normalized NPI.  Such a quantity gives a sense of how close a set of functions are to being sufficient statistics.  This topic will be discussed in \hyperref[sufficiency]{Section VI}. 

In \hyperref[math]{Section II} we will lay out some mathematical preliminaries and discuss the general transformations in statistical manifolds we are interested in.  Then in \hyperref[derive]{Section III}, we will state and discuss the design criteria used to derive the functional form of TC and the NPI in general.  In \hyperref[proof]{Section IV} we will complete the proof of the results from \hyperref[derive]{Section III}.  In \hyperref[npartite]{Section V} we discuss the $n$-partite (NPI) special cases of TC of which the bipartite case is the \hyperref[mi]{mutual information}.  In \hyperref[sufficiency]{Section VI} we will discuss sufficiency and its relation to the Neyman-Pearson lemma \cite{Neyman-Pearson}.  It should be noted that throughout this article we will be using a probabilistic framework in which $x \in \mathbf{X}$ denotes propositions of a probability distribution rather than a statistical framework in which $x$ denotes random numbers.

\section{Mathematical preliminaries}\label{math}
The arena of any inference task consists of two ingredients, the first of which is the subject matter, or what is often called \textit{the universe of discourse}.  This refers to the actual propositions that one is interested in making inferences about.  Propositions tend to come in two classes, either \textit{discrete} or \textit{continuous}.  Discrete proposition spaces will be denoted by calligraphic uppercase Latin letters, $\mathcal{X}$, and the individual propositions will be lowercase Latin letters $x_i \in \mathcal{X}$ indexed by some variable $i = \{1,\dots,|\mathcal{X}|\}$, where $|\mathcal{X}|$ is the number of distinct propositions in $\mathcal{X}$.  In this paper we will mostly work in the context of \textit{continuous} propositions whose spaces will be denoted by bold faced uppercase Latin letters, $\mathbf{X}$, and whose elements will simply be lowercase Latin letters with no indices, $x \in \mathbf{X}$.  Continuous proposition spaces have a much richer structure than discrete spaces (due to the existence of various differentiable structures, the ability to integrate, etc.) and help to generalize concepts such as \textit{relative entropy} and \textit{information geometry} \cite{Amari2,Cencov,Caticha_Book}\footnote{Common examples of discrete proposition spaces are the results of a coin flip or a toss of a die, while an example of a continuous proposition space is the position of a particle \cite{ED}.}.\\
\indent  The second ingredient that one needs to define for general inference tasks is the space of models, or the space of probability distributions which one wishes to assign to the underlying proposition space.  These spaces can often be given the structure of a manifold, which in the literature is called a \textit{statistical manifold} \cite{Caticha_Book}.  A statistical manifold $\mathbf{\Delta}$, is a manifold in which each point $\rho \in \mathbf{\Delta}$ is an entire probability distribution, i.e. $\mathbf{\Delta}$ is a space of maps from subsets of $\mathbf{X}$ to the interval $[0,1]$, $\rho:\mathcal{P}(\mathbf{X})\rightarrow[0,1]$.  The notation $\mathcal{P}(\mathbf{X})$ denotes the \textit{power set} of $\mathbf{X}$, which is the set of all subsets of $\mathbf{X}$, and has cardinality equal to $|\mathcal{P}(\mathbf{X})| = 2^{|\mathbf{X}|}$.\\
\indent  In the simplest cases, when the underlying propositions are discrete, the manifold is finite dimensional.  A common example that is used in the literature is the three-sided die, whose distribution is determined by three probability values $\rho = \{p_1,p_2,p_3\}$.  Due to positivity, $p_i \geq 0$, and the normalization constraint, $\sum_i p_i = 1$, the point $\rho$ lives in the $2$-simplex.  Likewise, a generic discrete statistical manifold with $n$ possible states is an $(n-1)$-simplex.  In the continuum limit, which is often the case explored in physics, the statistical manifold becomes infinite dimensional and is defined as\footnote{Throughout the rest of the paper, we use the Greek $\rho$ to represent a generic distribution in $\mathbf{\Delta}$, and we use the Latin $p(x)$ to refer to an individual density.},
\begin{equation}
\mathbf{\Delta} = \left\{p(x)\middle|\, p(x) \geq 0,\, \int dx\, p(x) = 1\right\}.\label{stat_man}
\end{equation}
When the statistical manifold is parameterized by the densities $p(x)$, the zeroes always lie on the boundary of the simplex.  In this representation the statistical manifolds have a trivial topology; they are all simply connected.  Without loss of generality, we assume that the statistical manifolds we are interested in can be represented as (\ref{stat_man}), so that $\mathbf{\Delta}$ is simply connected and does not contain any holes.  The space $\mathbf{\Delta}$ in this representation is also smooth.\\
\indent  The symbol $\rho$ defines what we call a \textit{state of knowledge} about the underlying propositions $\mathbf{X}$.  It is, in essence, the quantification of our \textit{degrees of belief} about each of the possible propositions $ x\in \mathbf{X}$ \cite{Cox}.  The \textit{correlations} present in any distribution $\rho$ necessarily depend on the conditional relationships between various propositions.  For instance, consider the binary case of just two proposition spaces $\mathbf{X}$ and $\mathbf{Y}$, so that the joint distribution factors,
\begin{equation}
p(x,y) = p(x)p(y|x) = p(y)p(x|y).\label{firstjoint}
\end{equation}
The correlations present in $p(x,y)$ will necessarily depend on the form of $p(x|y)$ and $p(y|x)$ since the conditional relationships tell us how one variable is statistically dependent on the other.  As we will see, the correlations defined in eq. (\ref{firstjoint}) are quantified by the \textit{mutual information}.  For situations of many variables however, the global correlations are defined by the \textit{total correlation}, which we will design first.  All other measures which break up the joint space into conditional distributions (including (\ref{firstjoint})) are special cases of the total correlation.  

\subsection{Some classes of inferential transformations}\label{class}

There are four main types of transformations we will consider that one can enact on a state of knowledge $\rho\stackrel{*}{\rightarrow}\rho'$. They are: \textbf{\textit{coordinate transformations, entropic updating\footnote{This of course includes Bayes rule as a special case \cite{Caticha_Giffin_1,Caticha_Giffin_2}}, marginalization}}, and  \textbf{\textit{products}}. This set of transformations is not necessarily exhaustive, but is sufficient for our discussion in this paper. We will indicate whether or not each of these types of transformations can presumably cause changes to the amount of global correlations, or not, by evaluating the response of the statistical manifold under these transformations. Our inability to describe \textit{how much} the amount of correlation changes under these transformations motivates the design of such an objective global quantifier.  

The types of transformations we will explore can be identified either with maps from a particular statistical manifold to itself, $\mathbf{\Delta}\rightarrow\mathbf{\Delta}$ (type \hyperref[coordtrans]{\textbf{I}}), to a subset of the original manifold $\mathbf{\Delta} \rightarrow \mathbf{\Delta}' \subseteq \mathbf{\Delta}$ (type \hyperref[entropicupdating]{\textbf{II}}), or from one statistical manifold to another, $\mathbf{\Delta}\rightarrow\mathbf{\Delta}'$ (type \hyperref[marginalization]{\textbf{III}} and \hyperref[products]{\textbf{IV}}).    

\subsubsection{Type I: Coordinate transformations}\label{coordtrans}
Type \textbf{I} transformations are coordinate transformations. A coordinate transformation $f:\mathbf{X}\rightarrow \mathbf{X}'$, is a special type of transformation of the proposition space $\mathbf{X}$ that respects certain properties.  It is essentially a continuous version of a reparameterization\footnote{A reparameterization is an isomorphism between discrete proposition spaces, $g:\mathcal{X}\rightarrow\mathcal{Y}$ which identifies for each proposition $x_i \in \mathcal{X}$, a unique proposition $y_i \in \mathcal{Y}$ so that the map $g$ is a bijection.}.  For one, each proposition $x \in \mathbf{X}$ must be identified with one and only one proposition $x'\in \mathbf{X}'$ and vice versa.  This means that coordinate transformations must be bijections on proposition space.  The reason for this is simply by design, i.e. we would like to study the transformations that leave the proposition space invariant.  A general transformation of type \textbf{I} on $\mathbf{\Delta}$ which takes $\mathbf{X}$ to $\mathbf{X}' = f(\mathbf{X})$, is met with the following transformation of the densities,
\begin{equation}
p(x) \xrightarrow{\mathbf{I}} p'(x') \quad \mathrm{where} \quad p(x)dx= p'(x')dx'.\label{i}
\end{equation}
Like we already mentioned, the coordinate transforming function $f:\mathbf{X}\rightarrow\mathbf{X}'$ must be a bijection in order for (\ref{i}) to hold, i.e. the map $f^{-1}:\mathbf{X}'\rightarrow\mathbf{X}$ is such that $f\circ f^{-1} = \mathrm{id}_{\mathbf{X}'}$ and $f^{-1}\circ f = \mathrm{id}_{\mathbf{X}}$.  While the densities $p(x)$ and $p'(x')$ are not necessarily equal, the probabilities defined in (\ref{i}) must be (according to the rules of probability theory, see the \hyperref[coordinate]{Appendix A}). This indicates that $\rho\stackrel{\textbf{I}}{\rightarrow}\rho'=\rho$ is in the same location in the statistical manifold. That is, the global state of knowledge has not changed -- what has changed is the way in which the local information in $\rho$ has been expressed, which must be invertible in general.\\  
\indent  While one could impose that the transformations $f$ be diffeomorphisms (i.e. smooth maps between $\mathbf{X}$ and $\mathbf{X}'$), it is not necessary that we restrict $f$ in this way.  Without loss of generality, we only assume that the bijections $f \in C^0(\mathbf{X})$ are continuous.  For discussions involving diffeomorphism invariance and statistical manifolds see the works of Amari \cite{Amari2}, Ay et. al. \cite{Ay} and Bauer et. al. \cite{Bauer}.\\
\indent  For a coordinate transformation (\ref{i}) involving two variables, $x \in \mathbf{X}$ and $y \in \mathbf{Y}$, we also have that type one transformations give,
\begin{equation}
p(x,y) \xrightarrow{\mathbf{I}} p'(x',y') \quad \mathrm{where}\quad p(x,y)dxdy = p'(x',y')dx'dy'.
\end{equation}
A few general properties of these type \textbf{I} transformations are as follows:  First, 
the density $p(x,y)$ is expressed in terms of the density $p'(x',y')$,
\begin{equation}
p(x,y) = p'(x',y')\gamma(x',y'),
\end{equation}
where $\gamma(x',y')$ is the determinant of the Jacobian \cite{Caticha_Book} that defines the transformation,
\begin{equation}
\gamma(x',y') = |\det\left(J(x',y')\right)|, \quad \mathrm{where}\quad J(x',y') = \begin{bmatrix}
\frac{\partial x'}{\partial x} & \frac{\partial x'}{\partial y} \\ \frac{\partial y'}{\partial x} & \frac{\partial y'}{\partial y}
\end{bmatrix}.
\end{equation}
For a finite number of variables $x = (x_1,\dots,x_N)$, the general type \textbf{I} transformations $p(x_1,\dots,x_N) \xrightarrow{\mathbf{I}} p'(x_1',\dots,x_N')$ are written,
\begin{equation}
p(x_1,\dots,x_N)\prod_{i=1}^Ndx_i = p'(x_1',\dots,x_N')\prod_{i=1}^Ndx_i',
\end{equation} 
and the Jacobian becomes,
\begin{equation}
J(x'_1,\dots,x'_N) = \begin{bmatrix}
\frac{\partial x_1'}{\partial x_1} &\cdots &\frac{\partial x_1'}{\partial x_N} \\ \vdots & \ddots & \vdots \\ \frac{\partial x_N'}{\partial x_1} & \cdots & \frac{\partial x_N'}{\partial x_N}
\end{bmatrix}.
\end{equation}
One can also express the density $p'(x')$ in terms of the original density $p(x)$ by using the inverse transform,
\begin{equation}
p'(x') = p(f^{-1}(x'))\gamma(x) = p(x)\gamma(x).
\end{equation}
In general, since coordinate transformations preserve the probabilities associated to a joint proposition space, they also preserve several structures derived from them.  One of these is the Fisher-Rao (information) metric \cite{Amari2,Bauer,Le}, which was proved by \v{C}encov \cite{Cencov} to be the unique metric on statistical manifolds that represents the fact that the points $\rho \in \mathbf{\Delta}$ are probability distributions and not structureless \cite{Caticha_Book}\footnote{For a summary of various derivations of the information metric, see \cite{Caticha_Book} section 7.4}.
\subsubsection{Split Invariant Coordinate Transformations}\label{splits}
%Throughout this paper, we will impose that all coordinate transformations which preserve the \textit{total correlation} in $\rho$
Consider a class of coordinate transformations that result in a diagonal Jacobian matrix, i.e.,
\begin{equation}
\gamma(x'_1,\dots,x'_N) = \prod_{i=1}^N\frac{\partial x'_i}{\partial x_i}.\label{diag}
\end{equation}
These transformations act within each of the variable spaces independently, and hence they are guaranteed to preserve the definition of the split between any $n$-partitions of the propositions, and because they are coordinate transformations, they are invertible and do not change our state of knowledge, $(\rho,\mathbf{X})\xrightarrow{\mathbf{Ia}}(\rho',\mathbf{X}') = (\rho,\mathbf{X})$.  We call such special types of transformations (\ref{diag}) \textit{split invariant coordinate transformations} and denote them as type \textbf{Ia}.  From (\ref{diag}), it is obvious that the marginal distributions of $\rho$ are preserved under split invariant coordinate transformations, 
\begin{equation}
p(x_i)dx_i = p'(x_i')dx_i'.\label{marginal}
\end{equation}
If one allows generic coordinate transformations of the joint space, then the marginal distributions may depend on variables outside of their original split. Thus, if one redefines the split after a coordinate transformation to new variables $\mathbf{X}\rightarrow\mathbf{X}'$, the original problem statement changes as to what variables we are considering correlations \emph{between} and thus eq. (\ref{marginal}) no longer holds. This is apparent in the case of two variables $(x,y)$, where $x' = f_{x'}(x,y)$, since,
\begin{equation}
dx' = df_{x'} = \frac{\partial f_{x'}}{\partial x}dx + \frac{\partial f_{x'}}{\partial y}dy,\label{coordtrans1}
\end{equation}  
which depends on $y$.  In the situation where $x$ and $y$ are independent, redefining the split after the coordinate transformation (\ref{coordtrans1}) breaks the original independence since the distribution that originally factors, $p(x,y) = p(x)p(y)$, would be made to have conditional dependence in the new coordinates, i.e. if $x' = f_{x'}(x,y)$ and $y' = f_{y'}(x,y)$, then,
\begin{equation}
p(x,y) = p(x)p(y) \xrightarrow{\mathbf{I}} p'(x',y') = p'(x')p'(y'|x').
\end{equation}
So, even though the above transformation satisfies (\ref{i}), this type of transformation may change the correlations in $\rho$ by allowing for the potential redefinition of the split $\mathbf{X}\rightarrow\mathbf{X}'$.  Hence, when designing our functional, we identify split invariant coordinate transformations as those which preserve correlations. These restricted coordinate transformations help isolate a single functional form for our global correlation quantifier.

\subsubsection{Type II: Entropic updating}\label{entropicupdating}
Type \textbf{II} transformations are those induced by updating \cite{Caticha_Book}, $\varphi\xrightarrow{*}\rho$ in which one maximizes the relative entropy,
\begin{equation}
S[\rho,\varphi] = -\int dx\, p(x)\log\frac{p(x)}{q(x)}\label{entropy},
\end{equation}
subject to constraints and relative to the prior, $q(x)$.  Constraints often come in the form of expectation values \cite{Jaynes1,Jaynes2,Jaynes_Book,Caticha_Book},
\begin{equation}
\langle f(x) \rangle = \int dx\, p(x)f(x) = \kappa.\label{constraints}
\end{equation} 
A special case of these transformations is Bayes' rule \cite{Caticha_Giffin_1,Caticha_Giffin_2},
\begin{equation}
p(x) \xrightarrow{\mathbf{II}} p'(x) \quad \mathrm{where} \quad p'(x) = p(x|\theta) = \frac{p(x)p(\theta|x)}{p(\theta)}\label{ii}.
\end{equation}
In (\ref{entropy}) and throughout the rest of the paper we will use $\log$ base $e$ (natural log) for all logarithms, although the results are perfectly well defined for any base (the quantities $S[\rho,\varphi]$ and $I[\rho,\mathbf{X}]$ will simply differ by an overall scale factor when using different bases).  Maximizing (\ref{entropy}) with respect to constraints such as (\ref{constraints}) induces a jump in the statistical manifold.  Type \textbf{II} transformations, while well defined, are not necessarily continuous, since in general one can map nearby points to disjoint subsets in $\mathbf{\Delta}$.  Type \textbf{II} transformations will also cause $\rho\stackrel{\textbf{II}}{\rightarrow}\rho'\neq\rho$ in general as it jumps within the statistical manifold. This means, because different $\rho$'s may have different correlations, that type \textbf{II} transformations can either increase, decrease, or leave the correlations invariant. 

\subsubsection{Type III: Marginalization}\label{marginalization}
Type \textbf{III} transformations are induced by marginalization,
\begin{equation}
p(x,y) \xrightarrow{\mathbf{III}} p(x) =  \int dy\, p(x,y)\label{iii},
\end{equation}
which is effectively a quotienting of the statistical manifold, $\mathbf{\Delta}(x) = \mathbf{\Delta}(x,y)/\sim_y$, i.e. for any point $p(x)$, we equivocate all values of $p(y|x)$.  Since the distribution $\rho$ changes under type \textbf{III} transformations, $\rho\xrightarrow{\mathbf{III}}\rho'$, the amount of correlations can change. 

\subsubsection{Type IV: Products}\label{products}
Type \textbf{IV} transformations are created by products,
\begin{equation}
p(x) \xrightarrow{\mathbf{IV}} p(x,y) = p(x)p(y|x)\label{iv},
\end{equation}
which are a kind of inverse transformation of type \textbf{III}, i.e. the set of propositions $\mathbf{X}$ becomes the product $\mathbf{X}\times\mathbf{Y}$.  There are many different situations that can arise from this type, a most trivial one being an embedding,
\begin{equation}
p(x) \xrightarrow{\mathbf{IVa}} p(x,y) = p(x)\delta(y - f(x))\label{iva},
\end{equation} 
which can be useful in many applications.  The function $\delta(\cdot)$ in the above equation is the \textit{Dirac delta function} \cite{Dirac} which has the following properties,
\begin{equation}
\delta(x) = \left\{\begin{matrix}
\infty & x = 0\\
0 & \mathrm{otherwise}\end{matrix} \right., \quad \mathrm{and} \quad \int dx\,\delta(x) = 1.
\end{equation}
We will denote such a transformation as type \textbf{IVa}.  Another trivial example of type \textbf{IV} is,
\begin{equation}
p(x) \xrightarrow{\mathbf{IVb}} p(x,y) = p(x)p(y)\label{ivb},   
\end{equation}
which we will call type \textbf{IVb}.  Like type \textbf{II}, generic transformations of type \textbf{IV} can potentially create correlations, since again we are changing the underlying distribution.
\subsection{Remarks on inferential transformations}
There are many practical applications in inference which make use of the above transformations by combining them in a particular order.  For example, in machine learning and dimensionality reduction, the task is often to find a low-dimensional representation of some proposition space $\mathbf{X}$, which is done by combining types \textbf{I},\textbf{III} and \textbf{IVa} in the order, $\rho\xrightarrow{\mathbf{IVa}}\rho'\xrightarrow{\mathbf{I}}\rho''\xrightarrow{\mathbf{III}}\rho'''$.  Neural networks are a prime example of this sequence of transformations \cite{Carrara_Ernst}.  Another example of \textbf{IV},\textbf{I},\textbf{III} transformations are \textit{convolutions} of probability distributions, which takes two proposition spaces and combines them into a new one \cite{CoverThomas_Book}.  

In \hyperref[consequences]{Appendix C} we discuss how our resulting design functionals behave under the aforementioned transformations. 

\section{Designing a global correlation quantifier}\label{derive}
In this section we seek to achieve our design goal for the special case of the \textit{total correlation},\\
\hrule
\vspace{1cm}

\textbf{Design Goal:} \textit{%Any global quantifier of correlations should be an increasing function of local correlations. 
	Given a space of $N$ variables $\mathbf{X} = \mathbf{X}_1\times\dots\times\mathbf{X}_N$ and a statistical manifold $\mathbf{\Delta}$, we seek to design a functional $I[\rho,\mathbf{X}]$\footnote{In Watanabe's paper \cite{Watanabe}, the notation for the \textit{total correlation} between a set of variables $\lambda$ is written as $C_{\mathrm{tot}}(\lambda) = \sum_{i}S(\lambda_i) - S(\lambda)$, where $S(\lambda_i)$ is the Shannon entropy of the subspace $\lambda_i\subseteq \lambda$.} which ranks distributions $\rho \in \mathbf{\Delta}$ according to their \textbf{total} amount of correlations.}
\vspace{1cm}
\hrule
\vspace{.3cm}

Unlike deriving a functional, designing a functional is done through the process of eliminative induction.  Derivations are simply a means of showing consistency with a proposed solution whereas design is much deeper.  In designing a functional, the solution is not assumed but rather achieved by specifying design criteria that restrict the functional form in a way that leads to a unique or optimal solution. One can then interpret the solution in terms of the original design goal. Thus, by looking at the ``nail", we design a ``hammer", and conclude that hammers are designed to knock in and remove nails. We will show that there are several paths to the solution of our design criteria, the proof of which is in \hyperref[proof]{Section IV}.  

Our design goal requires that $I[\rho,\mathbf{X}]$ be scalar valued such that we can rank the distributions $\rho$ according to their correlations. Considering a continuous space $\mathbf{X} = \mathbf{X}_1\times\dots\times\mathbf{X}_N$ of $N$ variables, the functional form of $I[\rho,\mathbf{X}]$ is the functional,
\begin{align}
I[\rho,\mathbf{X}]=I[p(x_1,\dots,x_N);p(x_1',\dots,x_N');\dots;\mathbf{X}]\label{I2},
\end{align}
which depends on each of the possible probability values for every $x \in \mathbf{X}$.
%where $\rho$ is expressed in all possible split invariant coordinates (parameterizations) in which the functional form of $I[\rho,\mathcal{X}]$ is invariant.  It is important to reiterate the discussion from \hyperref[coordtrans]{Section II} that for the correlations between each variable to remain fixed in the special case of the \textit{total correlation}, the coordinate representations that are allowed in (\ref{I}) are only those which are transformed within each variable so that the Jacobians for each densities in (\ref{I}) obey (\ref{diag}).

Given the types of transformations that may be enacted on $\rho$, we state the main guiding principle we will use to meet our design goal,\\ 
\hrule
\vspace{1cm}

\textbf{Principle of Constant Correlations (PCC):} \textit{The amount of correlations in $(\rho,\mathbf{X})$ should not change unless required by the transformation, $(\rho,\mathbf{X})\stackrel{*}{\rightarrow}(\rho',\mathbf{X}')$.}
\vspace{1cm}
\hrule
\vspace{.3cm}

While simple, the PCC is incredibly constraining. By stating when one should \emph{not} change the correlations, i.e. $I[\rho,\mathbf{X}]\stackrel{*}{\rightarrow} I[\rho',\mathbf{X}']=I[\rho,\mathbf{X}]$, it is operationally unique (i.e. that you don't do it) rather than stating how one is required to change them, $I[\rho,\mathbf{X}]\stackrel{*}{\rightarrow} I[\rho',\mathbf{X}']\neq I[\rho,\mathbf{X}]$, of which there are infinitely many choices. The PCC therefore imposes an element of objectivity into $I[\rho,\mathbf{X}]$.  If we are able to complete our design goal, then we will be able to uniquely quantify how transformations of type \textbf{I-IV} affect the amount of correlations in $\rho$.

The discussion of type $\mathbf{I}$ transformations indicate that split invariant coordinate transformations do not change $(\rho,\mathbf{X})$.  This is because we want to not only maintain the relationship among the joint distribution (\ref{i}), but also the relationships among the marginal spaces,
\begin{equation}
p(x_i)dx_i = p'(x_i')dx_i'.
\end{equation}
Only then are the relationships between the $n$-partitions guaranteed to remain fixed and hence the distribution $\rho$ remains in the same location in the statistical manifold. When a coordinate transformation of this type is made, because it does not change $(\rho,\mathbf{X})$, we are not explicitly required to change $I[\rho,\mathbf{X}]$, so by the PCC we impose that it does not.

The PCC together with the design goal implies that,\\
\hrule
\vspace{.5cm}
\begin{coro}[Split Coordinate Invariance]
	The coordinate systems within a particular split are no more informative about the amount of correlations than any other coordinate system for a given $\rho$\label{coro}.
\end{coro}    
\vspace{.5cm}
\hrule
\vspace{.3cm}
\noindent This expression is somewhat analogous to the statement that ``coordinates carry no information'', which is usually stated as a design criterion for relative entropy \cite{ShoreJohnson, Caticha_Entropy_1,Skilling}\footnote{This appears as axiom two in Shore and Johnson's derivation of relative entropy \cite{ShoreJohnson}, which is stated on page 27 as ``II. \textit{Invariance}: The choice of coordinate system should not matter.''  In Skilling's approach \cite{Skilling}, which was mainly concerned with image analysis, axiom two on page 177 is justified with the statement ``We expect the same answer when we solve the same problem in two different coordinate systems, in that the reconstructed images in the two systems should be related by the coordinate transformation.''  Finally, in Caticha's approach \cite{Caticha_Entropy_1}, the axiom of coordinate invariance is simply stated on page 4 as ``\textbf{Criterion 2: Coordinate invariance}. \textit{The system of coordinates carries no information}.''}.  

To specify the functional form of $I[\rho,\mathbf{X}]$ further, we will appeal to special cases in which it is apparent that the PCC should be imposed \cite{Skilling}. The first involves local, subdomain, transformations of $\rho$. If a subdomain of $\mathbf{X}$ is transformed then one may be required to change its amount of correlations by some specified amount. Through the PCC however, there is no explicit requirement to change the amount of correlations outside of this domain, hence we impose that those correlations outside are not changed. The second special case involves transformations of an independent subsystem. If a transformation is made on an independent subsystem then again by the PCC, because there is no explicit reason to change the amount of correlations in the other subsystem, we impose that they are not changed.  We denote these two types of transformation independences as our two design criteria (DC).

Surprisingly, the PCC and the DC are enough to find a general form for $I[\rho,\mathbf{X}]$ (up to an irrelevant scale constant).  As we previously stated, the first design criteria concerns local changes in the probability distribution $\rho$.\\  
\hrule
\vspace{0.5cm}
\begin{design}[Locality]
	Local transformations of $\rho$ contribute locally to the total amount of correlations.
	
\end{design}
\vspace{0.5cm}
\hrule
\vspace{.3cm}
The term \textit{locality} has been invoked to mean many different things in different fields (e.g. physics, statistics, etc.).  In this paper, as well as in \cite{ShoreJohnson,Caticha_Entropy_1,Skilling,Vanslette} and \cite{VansletteThesis}, the term \textit{local} refers to transformations which are constrained to act only within a particular subdomain $\mathcal{D} \subset \mathbf{X}$, i.e. the transformations of the probabilities are \textit{local} to $\mathcal{D}$ and do not affect probabilities outside of this domain.  Essentially, if new information does not require us to change the correlations in a particular subdomain $\mathcal{D} \subset \mathbf{X}$, then we don't change the probabilities over that subdomain.  While simple, this criterion is incredibly constraining and leads (\ref{I2}) to the functional form,
\begin{equation}
I[\rho,\mathbf{X}] \xrightarrow{DC1} \int dx \,F(p(x_1,\dots,x_N),x_1,\dots,x_N),\label{subsetindep}
\end{equation}
where $F$ is some undetermined function of the probabilities and possibly the coordinates.  We have used $dx = dx_1\dots dx_N$ to denote the measure for brevity.  To constrain $F$ further, we first use the corollary of split coordinate invariance (\ref{coro}) among the subspaces $\mathbf{X}_i \subset \mathbf{X}$ and then apply special cases of particular coordinate transformations.  This leads to the following functional form,
\begin{equation}
I[\rho,\mathbf{X}] \xrightarrow{PCC} \int dx\, p(x_1,\dots,x_N)\Phi\left(\frac{p(x_1,\dots,x_N)}{\prod_{i=1}^Np(x_i)} \right),\label{dc1coord}
\end{equation}
which demonstrates that the integrand is independent of the actual coordinates themselves.  Like coordinate invariance, the axiom DC1 also appears in the design derivations of relative entropy \cite{ShoreJohnson, Caticha_Entropy_1,Skilling,Vanslette,VansletteThesis}\footnote{In Shore and Johnson's approach to relative entropy \cite{ShoreJohnson}, axiom four is analogous to our locality criteria, which states on page 27 ``IV. \textit{Subset Independence}: It should not matter whether one treats an independent subset of system states in terms of a separate conditional density or in terms of the full system density.''  In Skilling's approach \cite{Skilling} locality appears as axiom one which, like Shore and Johnson's axioms, is called \textit{Subset Independence} and is justified with the following statement on page 175, ``Information about one domain should not affect the reconstruction in a different domain, provided there is no constraint directly linking the domains.''  In Caticha \cite{Caticha_Entropy_1} the axiom is also called \textit{Locality} and is written on page four as ``\textbf{Criterion 1: Locality}. \textit{Local information has local effects}.''  Finally, in Vanslette's work \cite{Vanslette,VansletteThesis}, the subset independence criteria is stated on page three as follows, ``Subdomain Independence:  When information is received about one set of propositions, it should not effect or change the state of knowledge (probability distribution) of the other propositions (else information was also received about them too).''}.\\
\indent  This leaves the function $\Phi$ to be determined, which can be done by imposing an additional design criteria.
\vspace{0.5cm}
\hrule
\vspace{0.5cm}
\begin{design}[Subsystem Independence]
	
	Transformations of $\rho$ in one independent subsystem can only change the amount of correlations in that subsystem.
	
\end{design}
\vspace{0.5cm}
\hrule
\vspace{.3cm}
\indent  The consequence of DC2 concerns independence among subspaces of $\mathbf{X}$.  Given two subsystems $(\mathbf{X}_1\times\mathbf{X}_2)\times(\mathbf{X}_3\times\mathbf{X}_4)\equiv \mathbf{X}_{12}\times\mathbf{X}_{34} = \mathbf{X}$ which are independent, the joint distribution factors,
\begin{equation}
p(x) = p(x_1,x_2)p(x_3,x_4) \quad \Leftrightarrow \quad \rho = \rho_{12}\rho_{34}.\label{subsets2}
\end{equation}
We will see that this leads to the global correlations being additive over each \textit{subsystem},
\begin{equation}
I[\rho,\mathbf{X}] = I[\rho_{12},\mathbf{X}_{12}] + I[\rho_{34},\mathbf{X}_{34}].
\end{equation}
Like locality (DC1), the design criteria concerning subsystem independence appears in all four approaches to relative entropy \cite{Caticha_Entropy_1,Skilling,ShoreJohnson,Vanslette,VansletteThesis}\footnote{In Shore and Johnson's approach \cite{ShoreJohnson}, axiom three concerns subsystem independence and is stated on page 27 as ``III. \textit{System Independence}: It should not matter whether one accounts for independent information about independent systems separately in terms of different densities or together in terms of a joint density.''  In Skillings approach \cite{Skilling}, the axiom concerning subsystem independence is given by axiom three on page 179 and provides the following comment on page 180 about its consequences ``This is the crucial axiom, which reduces S to the entropic form. The basic point is that when we seek an uncorrelated image from marginal data in two (or more) dimensions, we need to multiply the marginal distributions. On the other hand, the variational equation tells us to add constraints through their Lagrange multipliers. Hence the gradient $\delta S/\delta f$ must be the logarithm.''  In Caticha's design derivation \cite{Caticha_Entropy_1}, axiom three concerns subsystem independence and is written on page 5 as ``\textbf{Criterion 3: Independence}.  \textit{When systems are known to be independent it should not matter whether they are treated separately or jointly.}'' Finally, in Vanslette \cite{Vanslette,VansletteThesis} on page 3 we have ``Subsystem Independence:  When two systems are a priori believed to be independent and we only receive information about one, then the state of knowledge of the other system remains unchanged.''}; however, due to the difference in the design goal here, we end up imposing DC2 closer to that of the work of \cite{Vanslette,VansletteThesis} as we do not explicitly have the Lagrange multiplier structure in our design space.\\
\indent  Imposing DC2 leads to the final functional form of $I[\rho,\mathbf{X}]$,
\begin{equation}
I[\rho,\mathbf{X}] \xrightarrow{DC2} \int dx\, p(x_1,\dots,x_N)\log\frac{p(x_1,\dots,x_N)}{\prod_{i=1}^Np(x_i)},\label{dc2}
\end{equation}
with $p(x_i)$ being the split dependent marginals.
This functional is what is typically referred to as the \textit{total correlation}\footnote{The concept of total correlation TC was first introduced in Watanabe \cite{Watanabe} as a generalization to Shannon's definition of mutual information.  There are many practical applications of TC in the literature \cite{verSteeg1,verSteeg2,Gao1,verSteeg3}.} and is the unique result obtained from imposing the PCC and the corresponding design criteria.\\
\indent  As was mentioned throughout, these results are usually implemented as design criteria for relative entropy as well.  Shore and Johnson's approach \cite{ShoreJohnson} presents four axioms, of which III and IV are \textit{subsystem} and \textit{subset} independence.  \textit{Subset} independence in their framework corresponds to eq. (\ref{subsetindep}) and to the Locality axiom of Caticha \cite{Caticha_Entropy_1}.  It also appears as an axiom in the approaches by Skilling \cite{Skilling} and Vanslette \cite{Vanslette,VansletteThesis}.  Subsystem independence is given by axiom three in Caticha's work \cite{Caticha_Entropy_1}, axiom two in Vanslette's \cite{Vanslette,VansletteThesis} and axiom three in Skilling's \cite{Skilling}.  While coordinate invariance was invoked in the approaches by Skilling, Shore and Johnson and Caticha, it was later found to be unnecessary in the work by Vanslette \cite{Vanslette,VansletteThesis} who only required two axioms.  Likewise, we find that it is an obvious consequence of the PCC and does not need to be stated as a separate axiom in our derivation of the total correlation.

The work by Csisz\'{a}r \cite{Csiszar} provides a nice summary of the various axioms used by many authors (including Azc\'{e}l \cite{Azcel}, Shore and Johnson \cite{ShoreJohnson} and Jaynes \cite{Jaynes1}) in their definitions of information theoretic measures\footnote{A list is given on page 3 of \cite{Csiszar} which includes the following for conditions on an entropy function $H(P)$; (1) Positivity ($H(P) \geq 0$), (2) Expansibility (``expansion'' of $P$ by a new component equal to $0$ does not change $H(P)$, i.e. embedding in a space in which the probabilities of the new propositions are zero), (3) Symmetry ($H(P)$ is invariant under permutation of the probabilities), (4) Continuity ($H(P)$ is a continuous function of $P$), (5) Additivity ($H(P\times Q) = H(P) + H(Q)$), (6) Subadditivity ($H(X,Y) \leq H(X) + H(Y)$), (7) Strong additivity ($H(X,Y) = H(X) + H(Y|X)$), (8) Recursivity ($H(p_1,\dots,p_n) = H(p_1+p_2,p_3,\dots,p_n) + (p_1+p_2)H(\frac{p_1}{p_1+p_2},\frac{p_2}{p_1+p_2})$) and (9) Sum property ($H(P) = \sum_{i=1}^ng(p_i)$ for some function $g$).}.  One could associate the design criteria in this work to some of the common axioms enumerated in \cite{Csiszar}, although some of them will appear as consequences of imposing a specific design criterion, rather than as an ansatz.  For example, the \textit{strong additivity} condition (see sections \hyperref[margagain]{C.1.3} and \hyperref[prodagain]{C.1.4} of the appendix) is the result of imposing DC1 and DC2.  Likewise, the condition of \textit{positivity} (i.e. $I[\rho,\mathbf{X}] \geq 0$) and \textit{convexity} occurs as a consequence of the design goal, split coordinate invariance (SCI) and both of the design criteria.  \textit{Continuity} of $I[\rho,\mathbf{X}]$ with respect to $\rho$ is imposed through the design goal, and \textit{symmetry} is a consequence of DC1.  In summary, \textbf{Design Goal}$\rightarrow$\textit{continuity}, \textbf{DC1}$\rightarrow$\textit{symmetry}, (\textbf{DC1 + DC2})$\rightarrow$\textit{strong additivity}, (\textbf{Design Goal + SCI + DC1 + DC2})$\rightarrow$\textit{positivity + convexity}.  As was shown by Shannon \cite{Shannon_Book} and others \cite{Csiszar,Azcel}, various combinations of these axioms, as well as the ones mentioned in footnote 11, are enough to characterize entropic measures.

One could argue that we could have merely imposed these axioms at the beginning to achieve the functional $I[\rho,\mathbf{X}]$, rather than through the PCC and the corresponding design criteria.  The point of this article however, is to design the correlation functionals by using principles of inference, rather than imposing conditions on the functional directly\footnote{This point was also discussed in the conclusion section of Shore and Johnson \cite{ShoreJohnson} see page 33.}.  In this way, the resulting functionals are consequences of employing the inference framework, rather than postulated arbitrarily.  

One will recognize that the functional form of (\ref{dc2}) and the corresponding $n$-partite informations (\ref{npi}) have the form of a relative entropy.  Indeed, if one identifies the product marginal $\prod_{i=1}^Np(x_i)$ as a \textit{prior} distribution as in (\ref{entropy}), then it may be possible to find constraints (\ref{constraints}) which update the product marginal to the desired joint distribution $p(x)$.  One can then interpret the constraints as the generators of the correlations.  We leave the exploration of this topic to a future publication.

\section{Proof of the main result}\label{proof}
We will prove the results summarized in the previous section. Let a proposition of interest be represented by $x_i \in \mathcal{X}$ -- an $N$ dimensional coordinate $x_i = (x_{i}^1,\dots,x_{i}^N)$ that lives somewhere in the discrete and fixed proposition space $\mathcal{X} = \{x_1,\dots,x_i,\dots,x_{|\mathcal{X}|}\}$, with $|\mathcal{X}|$ being the cardinality of $\mathcal{X}$ (i.e. the number of possible combinations). The joint probability distribution at this generic location is $P(x_i)\equiv P(x_{i}^1,\dots,x_{i}^N)$ and the entire distribution $\rho$ is the set of joint probabilities defined over the space $\mathcal{X}$, i.e., $\rho \equiv \{P(x_1),\dots,P(x_{|\mathcal{X}|})\}\in \mathbf{\Delta}$. 

% Consider first the case where the proposition space is discrete $\mathcal{X} = \{x_1,\dots,x_{|\mathcal{X}|}\}$ and is $N$ dimensional so that a generic proposition $x_i \in \mathcal{X}$ represents values of each of the $N$ variables $x_i = \{x_{i}^1,\dots,x_{i}^N\}$.
\subsection{Locality - DC1}
We begin by imposing DC1 on $I[\rho,\mathcal{X}]$.  Consider changes in $\rho$ induced by some transformation $(*)$, where the change to the state of knowledge is,
\begin{equation}
\rho\xrightarrow{*}\rho' = \rho + \delta\rho\label{deltarho},
\end{equation}
for some arbitrary change $\delta\rho$ in $\mathbf{\Delta}$ that is required by some new information.  This implies that the global correlation function must also change according to (\ref{I2}),
\begin{equation}
I[\rho,\mathcal{X}] \xrightarrow{*} I[\rho',\mathcal{X}] = I[\rho,\mathcal{X}] + \delta I.
\end{equation}
where $\delta I$ is the change to $I[\rho,\mathcal{X}]$ induced by (\ref{deltarho}).  To impose DC1, consider that the new information requires us to change the distribution in one subdomain $\mathcal{D} \subset \mathcal{X}$, $\rho\rightarrow\rho'=\rho + \delta\rho_{\mathcal{D}}$, that may change the correlations, while leaving the probabilities in the complement domain fixed, $\delta\rho_{\bar{\mathcal{D}}}=0$.\footnote{The subdomain $\mathcal{D}$ and its compliment $\bar{\mathcal{D}}$ obey the relations, $\mathcal{D}\cap\bar{\mathcal{D}} = \emptyset$ and $\mathcal{D}\cup\bar{\mathcal{D}} = \mathcal{X}$. % The notation $p(x_i|\mathcal{D})$ represents the densities $p(x_i)$ which are in the subdomain $\mathcal{D}$. Therefore, the variation $\delta p(x_i|\bar{\mathcal{D}}) = 0$ means that the variations of densities outside of $\mathcal{D}$ are kept fixed.
} Let the subset of the propositions in $\mathcal{D}$ be relabeled as $\{x_1,\dots,x_d,\dots,x_{|\mathcal{D}|}\} \subseteq \{x_1,\dots,x_i,\dots,x_{|\mathcal{X}|}\}$.  Then the variations in $I[\rho,\mathcal{X}]$ with respect to the changes of $\rho$ in the subdomain $\mathcal{D}$ are, 
\begin{equation}
I[\rho,\mathcal{X}] \xrightarrow{*} I[\rho + \delta\rho_{\mathcal{D}},\mathcal{X}] \approx I[\rho,\mathcal{X}] + \sum_{d\in\mathcal{D}}\frac{\partial I[\rho,\mathcal{X}]}{\partial P(x_{d})}\delta P(x_{d}),\label{derives}
\end{equation}
for small changes $\delta P(x_{d})$. In general the derivatives in (\ref{derives}) are functions of the probabilities,
% \begin{equation}
% \frac{\partial I[\rho,\mathcal{X}]}{\partial p(x_{i}|\mathcal{D})} = f\left(p(x_1),...,p(x_{|\mathcal{X}|})\right)\label{f},
% \end{equation}
\begin{equation}
\frac{\partial I[\rho,\mathcal{X}]}{\partial P(x_{d})} = f_d\left(P(x_1),...,P(x_{|\mathcal{X}|}),x_d\right)\label{f},
\end{equation}
which could potentially depend on the entire distribution $\rho$ as well as the point $x_d \in \mathcal{X}$.  We impose DC1 by constraining (\ref{f}) to only depend on the probabilities within the subdomain $\mathcal{D}$ since the variation (\ref{f}) should not cause changes to the amount of correlations in the complement $\bar{\mathcal{D}}$, i.e.,
\begin{equation}
\frac{\partial I[\rho,\mathcal{X}]}{\partial P(x_d)} \xrightarrow{DC1} f_d\left(P(x_1),...,P(x_d),...,P(x_{|\mathcal{D}|}),x_d\right)\label{f2}.
\end{equation}
This condition must also hold for arbitrary choices of subdomains $\mathcal{D}$, thus by further imposing DC1 in the most restrictive case of local changes ($\mathcal{D} = x_d$),
\begin{equation}
\frac{\partial I[\rho,\mathcal{X}]}{\partial P(x_d)}\xrightarrow{DC1}f_d(P(x_d),x_d)\label{delta},
\end{equation}
guarantees that it will hold in the general case.  In this most restrictive case of local changes, the functional $I[\rho,\mathcal{X}]$ has vanishing mixed derivatives,
\begin{equation}
\frac{\partial^2I[\rho,\mathcal{X}]}{\partial P(x_i)P(x_j)} = 0, \quad \forall\, i \neq j.
\end{equation}
Integrating (\ref{delta}) leads to,
\begin{equation}
I[\rho,\mathcal{X}] = \sum_{i=1}^{|\mathcal{X}|}F_i(P(x_i),x_i) + \mathrm{const.},\label{36}
\end{equation}
where the $\{F_i\}$ are undetermined functions of the probabilities and the coordinates.  As this functional is designed for ranking, nothing prevents us from setting the irrelevant constant to zero, which we do. Extending to the continuum, we find eq. (\ref{subsetindep}),
\begin{equation}
I[\rho,\mathbf{X}]= \int dx\, F\left(p(x),x\right)\label{sum},
\end{equation}
where for brevity we have also condensed the notation for the continuous $N$ dimensional variables $x = \{x_1,\dots,x_N\}$.  It should be noted that $ F(p(x),x)$ has the capacity to express a large variety of potential measures of correlation including Pearson's \cite{Pearson} and Szekely's \cite{Szekely} correlation coefficients. Our new objective is to use eliminative induction until only a unique functional form for $F$ remains.

\subsubsection{Split coordinate invariance -- PCC}
\indent  The PCC and the corollary (\ref{coro}) state that $I[\rho,\mathbf{X}]$, and thus $F\left(p(x),x\right)$, should be independent of transformations that keep $(\rho,\mathbf{X})\xrightarrow{*} (\rho',\mathbf{X}')=(\rho,\mathbf{X})$ fixed. As discussed, split invariant coordinate transformations (\ref{diag}) satisfy this property.  We will further restrict the functional $I[\rho,\mathbf{X}]$ so that it obeys these types of transformations. 

We can always rewrite the expression (\ref{sum}) by introducing densities $m(x)$ and $p(x)$ so that,
\begin{equation}
I[\rho,\mathbf{X}] = \int dx\, p(x)\frac{1}{p(x)}F\left(\frac{p(x)}{m(x)}m(x),x\right).
\end{equation}
Then, instead of dealing with the function $F$ directly, we can instead deal with a new definition $\Phi$,
\begin{equation}
I[\rho,\mathbf{X}] = \int dx\, p(x)\Phi\left(\frac{p(x)}{m(x)},p(x),m(x),x\right),
\end{equation}
where $\Phi$ is defined as,
\begin{equation}
\Phi\left(\frac{p(x)}{m(x)},p(x),m(x),x\right) \stackrel{\mathrm{def}}{=} \frac{1}{p(x)}F\left(\frac{p(x)}{m(x)}m(x),x\right)\label{definition}.
\end{equation}
Now we further restrict the functional form of $\Phi$ by appealing to the PCC.  Consider the functional $I[\rho,\mathbf{X}]$ under a split invariant coordinate transformation, 
\begin{align}
(x_1,\dots,x_N) \rightarrow (x'_1,\dots,x'_N) \quad \Rightarrow \quad m(x)dx &= m'(x')dx',\nonumber\\
\quad \mathrm{and}\quad p(x)dx &= p'(x')dx',\label{transforms}
\end{align}
which amounts to sending $\Phi$ to,
\begin{equation}
\Phi\left(\frac{p'(x')}{m'(x')},p'(x'),m'(x'),x'\right) = \Phi\left(\frac{p(x)}{m(x)},\gamma(x)p(x),\gamma(x)m(x),x'\right),
\end{equation}
where $\gamma(x) = \prod_{i}^N\gamma(x_i)$ is the Jacobian for the transformation from $(x_1,\dots,x_N)$ to $(f_1(x_1),\dots,f_N(x_N))$.  
Consider the special case in which the Jacobian $\gamma(x) = 1$.  Then due to the PCC we must have,
\begin{equation}
\Phi\left(\frac{p(x)}{m(x)},p(x),m(x),x\right) = \Phi\left(\frac{p(x)}{m(x)},p(x),m(x),x'\right).
\end{equation}
However this would suggest that $I[\rho,\mathbf{X}] \xrightarrow{\mathbf{Ia}}I[\rho',\mathbf{X}'] \neq I[\rho,\mathbf{X}]$ since correlations could be changed under the influence of the new variables $x' \in \mathbf{X}'$.  Thus in order to maintain the global correlations the function $\Phi$ must be independent of the coordinates,
\begin{equation}
\Phi \xrightarrow{PCC} \Phi\left(\frac{p(x)}{m(x)},p(x),m(x)\right).
\end{equation}
To constrain the form of $\Phi$ further, we can again appeal to split coordinate invariance but now with arbitrary Jacobian $\gamma(x) \neq 1$, which causes $\Phi$ to transform as,
\begin{equation}
\Phi\left(\frac{p(x)}{m(x)},p(x),m(x)\right) = \Phi\left(\frac{p(x)}{m(x)},\gamma(x)p(x),\gamma(x)m(x)\right).
\end{equation}
But this must hold for arbitrary split invariant coordinate transformations, for when the Jacobian factor $\gamma(x) \neq 1$.  Hence, the function $\Phi$ must also be independent of the second and third argument,
\begin{equation}
\Phi \xrightarrow{PCC} \Phi\left(\frac{p(x)}{m(x)}\right).
\end{equation}
We then have that the split coordinate invariance suggested by the PCC together with DC1 gives,
\begin{equation}
I[\rho,\mathbf{X}] = \int dx\, p(x)\Phi\left(\frac{p(x)}{m(x)}\right)\label{step2}.
\end{equation}
This is similar to the steps found in the relative entropy derivation \cite{ShoreJohnson,Caticha_Entropy_1}, but differs from the steps in \cite{Vanslette,VansletteThesis}. 
\subsubsection{$I_{min}$ -- Design Goal and PCC}
Split coordinate invariance, as realized in eq. (\ref{step2}), provides an even stronger restriction on $I[\rho,\mathbf{X}]$ which we can find by appealing to a special case.  Since all distributions with the same correlations should have the same value of $I[\rho,\mathbf{X}]$ by the Design Goal and PCC, then all independent joint distributions $\varphi$ will also have the same value, which by design takes a unique minimum value,
\begin{equation}
p(x) = \prod_{i=1}^Np(x_i) \Longrightarrow I[\varphi,\mathbf{X}] = I_{\mathrm{min}}.\label{min}
\end{equation}
Requiring independent joint distributions $\varphi$ return a unique minimum $I[\varphi,\mathbf{X}] = I_{\mathrm{min}}$ is similar to imposing a \textit{positivity} condition on $I[\rho,\mathbf{X}]$ \cite{Csiszar}.  We will find however, that positivity only arises once DC2 has been taken into account.  Here, $I_{\mathrm{min}}$ could be any value, so long as when one introduces correlations, $\varphi \xrightarrow{*}\rho$ the value of $I[\rho,\mathbf{X}]$ always increases from $I_{\mathrm{min}}$.  This condition could also be imposed as a general \textit{convexity} property of $I[\rho,\mathbf{X}]$, however this is already required by the \hyperref[derive]{design goal} and does not require an additional axiom. 

Inserting (\ref{min}) into (\ref{step2}) we find,
\begin{equation}
I[\rho,\mathbf{X}] = \int dx\, \Big(\prod_{i=1}^Np(x_i)\Big)\Phi\left(\frac{\prod_{i=1}^Np(x_i)}{m(x)}\right) = I_{\mathrm{min}}\label{ind}.
\end{equation}
But this expression must be independent of the underlying distribution $p(x)=\prod_{i=1}^Np(x_i)$, since all independent distributions, regardless of the joint space $\mathbf{X}$, must give the same value $I_{\mathrm{min}}$.  Thus we conclude that the density $m(x)$ must be the product marginal $m(x) = \prod_{i=1}^Np(x_i)$,
\begin{equation}
p(x_i) = \int d\bar{x}_i\, p(x), \qquad \mathrm{where} \quad d\bar{x}_i = \prod_{k\neq i}dx_k,
\end{equation}
so it is guaranteed that,
\begin{equation}
I_{\mathrm{min}} = \int dx\, p(x)\Phi(1)=\Phi(1) = \mathrm{const.}\label{constphi}
\end{equation}
Thus, by design, expression (\ref{step2}) becomes (\ref{dc1coord}),
\begin{equation}
I[\rho,\mathbf{X}] \xrightarrow{PCC} \int dx\, p(x)\Phi\left(\frac{p(x)}{\prod_{i=1}^Np(x_i)}\right)\label{step9}.
\end{equation}

\subsection{Subsystem Independence -- DC2}
In the following subsections we will consider two approaches for imposing subsystem independence via the PCC and DC2. Both lead to identical functional expressions for $I[\rho,\mathbf{X}]$. The analytic approach assumes the functional form of $\Phi$ may be expressed as a Taylor series. The algebraic approach reaches the same conclusion without this assumption.
\subsubsection{Analytical Approach}
Let us assume that the function $\Phi$ is analytic, so that it can be Taylor expanded.  Since the argument, $p(x)/\prod_{i=1}^Np(x_i)$ is defined over $[0,\infty)$, we can consider the expansion over some open set of $[0,\infty)$ for any particular value $p_0(x)/\prod_{i=1}^Np_0(x_{i})$ as,
\begin{equation}
\Phi\left(\frac{p(x)}{\prod_{i=1}^Np(x_i)}\right) = \sum_{n=0}^{\infty}\tilde{\Phi}_n\left(\frac{p(x)}{\prod_{i=1}^Np(x_i)} - \frac{p_0(x)}{\prod_{i=1}^Np_0(x_i)}\right)^n,\label{series}
\end{equation}
where $\tilde{\Phi}_n$ are real coefficients.  For $p(x)/\prod_{i=1}^Np(x_i)$ in the neighborhood of $p_0(x)/\prod_{i=1}^Np_0(x_i)$, the series (\ref{series}) converges to $\Phi\left[p(x)/\prod_{i=1}^Np(x_i)\right]$.
The Taylor expansion of $\Phi\left[\frac{p(x)}{\prod_{i=1}^Np(x_i)}\right]$ about $p(x)$ when its propositions are nearly independent, i.e. $p(x) \approx \prod_{i=1}^Np(x_i)$, is
\begin{equation}
\Phi=\sum_{n=0}^{\infty}\frac{1}{n!}\Phi^{(n)}\left[\frac{p(x)}{\prod_{i=1}^Np(x_i)}\right]\Big(\frac{p(x)}{\prod_{i=1}^Np(x_i)}-1\Big)^n,
\end{equation}
where the upper index $(n)$ denotes the $n$th-derivative,
\begin{equation}
\Phi^{(n)}\left[\frac{p(x)}{\prod_{i=1}^Np(x_i)}\right]=\Big(\frac{\delta}{\delta p(x)}\Big)^{(n)}\Phi\left[\frac{p(x)}{\prod_{i=1}^Np(x_i)}\right]\Big|_{p(x)=\prod_{i=1}^Np(x_i)}.
\end{equation}
The 0th term is $\Phi^{(0)}\left[\frac{p(x)}{\prod_{i=1}^Np(x_i)}\right]=\Phi[1]=\Phi_{min}$ by definition of the design goal, which leaves,
\begin{equation}
\Phi^+=\sum_{n=1}^{\infty}\frac{1}{n!}\Phi^{(n)}\left[\frac{p(x)}{\prod_{i=1}^Np(x_i)}\right]\Big(\frac{p(x)}{\prod_{i=1}^Np(x_i)}-1\Big)^n,
\end{equation}
where the $+$ in $\Phi^+$ refers to $(n) > 0$.

Consider the independent subsystem special case in which $p(x)$ is factorizable into $p(x)=p(x_1,x_2)p(x_3,x_4)$, for all $x \in \mathbf{X}$. We can represent $\Phi^+$ with an analogous two-dimensional Taylor expansion in $p(x_1,x_2)$ and $p(x_3,x_4)$, which is,
\begin{align}
\Phi^+=&\sum_{n_1=1}^{\infty}\frac{1}{n_1!}\Phi^{(n_1)}\left[\frac{p(x_1,x_2)}{p(x_1)p(x_2)}\right]\Big(\frac{p(x_1,x_2)}{p(x_1)p(x_2)}-1\Big)^{n_1}\nonumber\\
+&\sum_{n_2=1}^{\infty}\frac{1}{n_2!}\Phi^{(n_2)}\left[\frac{p(x_3,x_4)}{p(x_3)p(x_4)}\right]\Big(\frac{p(x_3,x_4)}{p(x_3)p(x_4)}-1\Big)^{n_2}\nonumber\\
+&\Bigg\{\sum_{\mathclap{\substack{n_1=1 \\ n_2=1}}}^{\infty}\frac{1}{n_1!n_2!}\Phi^{(n_1,n_2)}\left[\frac{p(x)}{\prod_{i=1}^Np(x_i)}\right]\nonumber\\
&\times\Big(\frac{p(x_1,x_2)}{p(x_1)p(x_2)}-1\Big)^{n_1}\Big(\frac{p(x_3,x_4)}{p(x_3)p(x_4)}-1\Big)^{n_2}\Bigg\},
\end{align}
where the mixed derivative term is,
\begin{align}
\Phi^{(n_1,n_2)}\left[\frac{p(x)}{\prod_{i=1}^Np(x_i)}\right] =& \left(\frac{\delta}{\delta p(x_1,x_2)}\right)^{(n_1)}\left(\frac{\delta}{\delta p(x_3,x_4)}\right)^{(n_2)}\nonumber\\
&\times\Phi\left[\frac{p(x)}{\prod_{i=1}^Np(x_i)}\right]\Big|_{p(x)=\prod_{i=1}^Np(x_i)}.
\end{align}
Since transformations of one independent subsystem, $\rho_{12}\xrightarrow{*}\rho'_{12}$ or $\rho_{34}\xrightarrow{*}\rho'_{34}$, must leave the other invariant by the PCC and subsystem independence, then DC2 requires that the mixed derivatives should necessarily be set to zero, $\Phi^{(n_1,n_2)}\left[\frac{p(x_1,x_2)p(x_3,x_4)}{\prod_{i=1}^Np(x_i)}\right]=0$. This gives a functional equation for $\Phi^+$,
\begin{align}
\Phi^+=&\sum_{n_1=1}^{\infty}\frac{1}{n_1!}\Phi^{(n_1)}\left[\frac{p(x_1,x_2)}{p(x_1)p(x_2)}\right]\Big(\frac{p(x_1,x_2)}{p(x_1)p(x_2)}-1\Big)^{n_1}\nonumber\\
+&\sum_{n_2=1}^{\infty}\frac{1}{n_2!}\Phi^{(n_2)}\left[\frac{p(x_3,x_4)}{p(x_3)p(x_4)}\right]\Big(\frac{p(x_3,x_4)}{p(x_3)p(x_4)}-1\Big)^{n_2} = \Phi^+_1 + \Phi^+_2,
\end{align}
where $\Phi_1^+$ corresponds to the terms involving $\mathbf{X}_1$ and $\mathbf{X}_2$ and $\Phi^+_2$ corresponds to the terms involving $\mathbf{X}_3$ and $\mathbf{X}_4$.  Including $\Phi_{min}=\Phi(1)$ from the $n_1=0$ and $n_2=0$ cases we have in total that,
\begin{equation}
\Phi = 2\Phi_{min} + \Phi^+_1 + \Phi^+_2.
\end{equation}
To determine the solution of this equation we can appeal to the special case in which both subsystems are independent, $p(x_1,x_2) = p(x_1)p(x_2)$, and $p(x_3,x_4) = p(x_3)p(x_4)$ which amounts to,
\begin{equation}
\Phi = \Phi_{\mathrm{min}} = 2\Phi_{\mathrm{min}},
\end{equation}
which means that either $\Phi_{\mathrm{min}} = 0$ or $\Phi_{\mathrm{min}} = \pm\infty$, however the latter two solutions are ruled out by the design goal since setting the minimum to $+\infty$ makes no sense, and setting it to $-\infty$ does not allow for ranking as it implies $\Phi=-\infty$ for all finite values of $\Phi^+$, which would violate the Design Goal. Further, $\Phi_{\mathrm{min}}=-\infty$ would imply that the minimum would not be a well defined constant number $\Phi$, which violates (\ref{constphi}). Thus, by eliminative induction and following our design method, it follows that $\Phi_{\mathrm{min}}=\Phi(1)$ must equal $0$.

The general equation for $\Phi$ having two independent subsystems $\rho=\rho_{12}\rho_{34}$ is,
\begin{equation}
\Phi[\rho_{12}\rho_{34}] = \Phi_1^+[\rho_{12}] + \Phi_2^+[\rho_{34}],
\end{equation}
or with the arguments,
\begin{equation}
\Phi\left[\frac{p(x_1,x_2)p(x_3,x_4)}{p(x_1)p(x_2)p(x_3)p(x_4)}\right] = \Phi_{1}^+\left[\frac{p(x_1,x_2)}{p(x_1)p(x_2)}\right] + \Phi_{2}^+\left[\frac{p(x_3,x_4)}{p(x_3)p(x_4)}\right].
\end{equation}
If subsystem $\rho_{34}=\rho_3\rho_4$ is itself independent, it implies 
\begin{equation}
\Phi\left[\frac{p(x_1,x_2)}{p(x_1)p(x_2)}\cdot 1\right] = \Phi_1^+\left[\frac{p(x_1,x_2)}{p(x_1)p(x_2)}\right],
\end{equation}
but due to commutativity, this is also, 
\begin{equation}
\Phi\left[1\cdot \frac{p(x_1,x_2)}{p(x_1)p(x_2)}\right] = \Phi_2^+\left[\frac{p(x_1,x_2)}{p(x_1)p(x_2)}\right].
\end{equation}
This implies the functional form of $\Phi$ does not have dependence on the particular subsystem $\Phi=\Phi_1^+=\Phi_2^+$ in general. This gives the following functional equation for $\Phi$,
\begin{equation}
\Phi\left[\frac{p(x_1,x_2)p(x_3,x_4)}{p(x_1)p(x_2)p(x_3)p(x_4)}\right] = \Phi\left[\frac{p(x_1,x_2)}{p(x_1)p(x_2)}\right] + \Phi\left[\frac{p(x_3,x_4)}{p(x_3)p(x_4)}\right].\label{log2}
\end{equation}
The solution to this functional equation is the log,
\begin{equation}
\Phi[z]=A\log(z),\label{log}
\end{equation}
where $A$ is an arbitrary constant.  Setting $A = 1$, the global correlation functional increases in the amount of correlations,
\begin{equation}
I[\rho,\mathbf{X}] = \int dx\, p(x)\log\frac{p(x)}{\prod_{i=1}^Np(x_i)}\label{MI},
\end{equation}
which is (\ref{dc2}).

The result in (\ref{log2}) could be imposed as an \textit{additivity} condition on the functional $I[\rho,\mathbf{X}] = I[\rho_{12},\mathbf{X}_{12}] + I[\rho_{34},X_{34}]$ \cite{Csiszar}.  In general however, the correlation functional $I[\rho,\mathbf{X}]$ obeys the stricter \textit{strong additivity} condition, which we have no reason a priori to impose.  Here, the \textit{strong additivity} condition is instead an end result, realized as a consequence of imposing the PCC through the various design criteria.

\subsubsection{Algebraic Approach}
Here we present an alternative algebraic approach to imposing DC2.  Consider the case in which subsystem two is independent, $\rho_{34} \rightarrow \varphi_{34} = p(x_3,x_4) = p(x_3)p(x_4)$\footnote{The notation $\varphi$ in place of the usual $\rho$ for a distribution is meant to represent independence among it's subsystems.}, and $\rho=\rho_{12}\varphi_{34}$. This special case is,
\begin{align}
I[\rho_{12}\varphi_{34},\mathbf{X}] &= \int dx\, p(x_1,x_2)p(x_3,x_4) \Phi\left[\frac{p(x_1,x_2)p(x_3,x_4)}{p(x_1)p(x_2)p(x_3)p(x_4)}\right]\nonumber\\
&= \int dx_1dx_2\, p(x_1,x_2)\Phi\left[\frac{p(x_1,x_2)}{p(x_1)p(x_2)}\right]=I[\rho_{12},\mathbf{X}_{12}],\label{I_1}
\end{align}
which holds for all product forms of $\varphi_{34}$ that have no correlations and for all possible transformations of $\rho_{12}\xrightarrow{*} \rho_{12}'$.joint

Alternatively, we could have considered the situation in which subsystem one is independent, $\rho_{12} \rightarrow \varphi_{12} = p(x_1,x_2) = p(x_1)p(x_2)$. Analogously, this case implies,
\begin{equation}
I[\varphi_{12}\rho_{34},\mathbf{X}] =I[\rho_{34},\mathbf{X}_{34}],
\end{equation}
which holds for all product forms of $\varphi_{12}$ that have no correlations and for all possible transformations of $\rho_{34}\xrightarrow{*} \rho_{34}'$. 

The consequence of these considerations is that, in principle, we have isolated the amount of correlations of either system. Imposing DC2 is requiring that the amount of correlations in either subsystem cannot be affected by changes in correlations in the other. This implies that for general $\rho=\rho_1\rho_2$,
\begin{equation}
I[\rho_1\rho_2,\mathbf{X}] = G[I[\rho_1,\mathbf{X}_1],I[\rho_2,\mathbf{X}_2]].\label{F}
\end{equation}
%or due to what is known about the functional form of $I[\rho]$, that,
%\begin{equation}
%\Phi\left[\frac{\rho_1\rho_2}{\varphi_1\varphi_2}\right] = G\Big[\Phi\left[\frac{\rho_1}{\varphi_1}\right],\Phi\left[\frac{\rho_2}{\varphi_2}\right]\Big].\label{functional}
%\end{equation}
%This is a functional equation over the functional $I$ which we can solve by appealing to special cases.  Consider first relabeling $f = \rho_1/\varphi_1$ and $g = \rho_2/\varphi_2$ so that (\ref{functional}) is written,
%\begin{equation}
%\Phi\left[fg\right]=G\Big[\Phi\left[f\right],\Phi\left[ g\right]\Big].
%\end{equation}
Consider a variation of $\rho_1$ where $\rho_2$ is held fixed, which induces a change in $I[\rho_1\rho_2,\mathbf{X}]$,
\begin{equation}
\delta I[\rho_1\rho_2,\mathbf{X}]\Big|_{\rho_2} = \frac{\delta I[\rho_1\rho_2,\mathbf{X}]}{\delta I[\rho_1,\mathbf{X}_1]}\frac{\delta I[\rho_1,\mathbf{X}_1]}{\delta \rho_1}\delta\rho_1.\label{variation}
\end{equation}
Now consider a variation of $\rho_1$ at any other value of the second subsystem, $\rho_2\xrightarrow{*}\rho_2'$. This is,
\begin{equation}
\delta I[\rho_1\rho_2',\mathbf{X}]\Big|_{\rho_2'} = \frac{\delta I[\rho_1\rho_2',\mathbf{X}]}{\delta I[\rho_1,\mathbf{X}_1]}\frac{\delta I[\rho_1,\mathbf{X}_1]}{\delta \rho_1}\delta\rho_1.\label{variation2}
\end{equation}
It follows from DC2 that transformations in one independent subsystem should not change the amount of correlations in another independent subsystem due to the PCC.  However, for the same $\delta \rho_1$, the current functional form (\ref{variation}) allows for $\delta I\left[\rho_1\rho_2,\mathbf{X}\right]/\delta \rho_1$ at one value of $\rho_2$ to differ from $\delta I\left[\rho_1\rho_2',\mathbf{X}\right]/\delta \rho_1$ at another, which implies that the amount of correlations induced by the change $\delta \rho_1$ depends on the value of $\rho_2$. Imposing DC2 is therefore enforcing that functionally the amount of change in the correlations satisfies
\begin{equation}
\frac{\delta I\left[\rho_1\rho_2,\mathbf{X}\right]}{\delta  I\left[\rho_1,\mathbf{X}_1\right]}=\frac{\delta I\left[\rho_1\rho_2',\mathbf{X}\right]}{\delta I\left[\rho_1,\mathbf{X}_1\right]},
\end{equation}
for any value of $\rho_2'$, i.e. that the variations must be independent too. This similarly goes for variations with respect to $\rho_2$ where $\rho_1$ is kept fixed, which implies that (\ref{F}) must be linear since,
\begin{equation}
\frac{\delta^2 I\left[\rho_1\rho_2,\mathbf{X}\right]}{\delta I\left[\rho_1,\mathbf{X}_1\right]\delta I\left[\rho_2,\mathbf{X}_2\right]}=0.
\end{equation}
The general solution to this differential equation is,
\begin{equation}
I\left[\rho_1\rho_2,\mathbf{X}\right]=a I\left[\rho_1,\mathbf{X}_1\right]+b I\left[\rho_2,\mathbf{X}_2\right]+c.
\end{equation}joint
We now seek the constants $a,b,c$. Commutativity, $I\left[\rho_1\rho_2,\mathbf{X}\right]=I\left[\rho_2\rho_1,\mathbf{X}\right]$, implies that $a=b$,
\begin{align}
I[\rho_1\rho_2,\mathbf{X}] &= a(I[\rho_1,\mathbf{X}_1]+I[\rho_2,\mathbf{X}_2])+c.
\end{align}
Because $I_{min}=\Phi[1]=\Phi[1^N]$, for $N$ independent subsystems we find,
\begin{align}
I[\varphi_1...\varphi_N,\mathbf{X}] = I_{min}= NaI_{min}+(N-1)c,
\end{align}
and therefore the constant $c$ must satisfy,
\begin{align}
c= \frac{(1-Na)I_{min}}{(N-1)}.
\end{align}
Because $a,c$, and $I_{min}$ are all constants they should not depend on the number of independent subsystems $N$. Thus, for another distribution $\rho$ which contains $M\neq N$ independent subsystems,
\begin{align}
c= \frac{(1 - Na)I_{min}}{(N-1)}= \frac{(1-Ma)I_{min}}{(M-1)},
\end{align}
which implies $N=M$, which can't be realized by definition. This implies the only solution is $I_{min}=c=0$, which is in agreement with the analytic approach. One then uses (\ref{I_1}),
\begin{align}
I[\rho_1\varphi_2,\mathbf{X}] = aI[\rho_1,\mathbf{X}_1]= I[\rho_1,\mathbf{X}_1],
\end{align}
and finds $a=1$.  This gives a functional equation for $\Phi$,
\begin{align}
\Phi[\rho_{12}\rho_{34}]= \Phi[\rho_{12}] + \Phi[\rho_{34}].
\end{align}
At this point the solution follows from eq. (\ref{log}) so that $I[\rho]$ is (\ref{dc2}),
\begin{equation}
I[\rho,\mathbf{X}] = \int dx\, p(x)\log\frac{p(x)}{\prod_{i=1}^Np(x_i)}\label{MI2}.
\end{equation}

\section{The $n$-partite special cases}\label{npartite}
In the previous sections of the article, we designed an expression that quantifies the global correlations present within an entire probability distribution and found this quantity to be identical to the total correlation (TC).  Now we would like to discuss partial cases of the above in which one does not consider the information shared by the entire set of variables $\mathbf{X}$, but only information shared \textit{across} particular subsets of variables in $\mathbf{X}$.  These types of special cases of TC measure the $n$-partite correlations present for a given distribution $\rho$.  We call such functionals an \textit{$n$-partite information}, or NPI.\\
\indent  Given a set of $N$ variables in proposition space, $\mathbf{X} = \mathbf{X}_1\times\dots\times\mathbf{X}_N$, an $n$-partite subset of $\mathbf{X}$ consists of $n\leq N$ subspaces $\{\mathbf{X}^{(k)}\}_n\equiv \{\mathbf{X}^{(1)},...,\mathbf{X}^{(k)},...,\mathbf{X}^{(n)}\}$ which have the following collectively exhaustive and mutually exclusive properties,
\begin{equation}
\mathbf{X}^{(1)}\times...\times\mathbf{X}^{(n)} = \mathbf{X} \quad \mathrm{and}\quad \mathbf{X}^{(k)}\cap\mathbf{X}^{(j)} = \emptyset, \,\forall\, k \neq j.
\end{equation}  
The special case of (\ref{MI2}) for any $n$-partite splitting will be called the \textit{$n$-partite information} and will be denoted by $I[\rho,\mathbf{X}^{(1)};\dots;\mathbf{X}^{(n)}]$ with $(n-1)$ semi-colons separating the partitions.  The largest number $n$ that one can form for any variable set $\mathbf{X}$ is simply the number of variables present in $\mathbf{X}$ and for this largest set the $n$-partite information coincides with the total correlation,
\begin{equation}
I[\rho,\mathbf{X}_1;\dots;\mathbf{X}_N] \stackrel{\mathrm{def}}{=} I[\rho,\mathbf{X}].
\end{equation}
\indent  Each of the $n$-partite informations can be derived in a manner similar to the total correlation, except where the density $m(x)$ in step (\ref{step9}) is replaced with the appropriate independent density associated to the $n$-partite system, i.e.,
\begin{equation}
m(x) \xrightarrow{\mathrm{n-partite}} \prod_{k=1}^np(x^{(k)}), \quad x^{(k)} \in \mathbf{X}^{(k)}.
\end{equation} 
Thus, the split invariant coordinate transformation (\ref{diag}) becomes one in which each of the partitions in variable space gives an overall block diagonal Jacobian,\footnote{In the simplest case, for $N$ dimensions and $n=2$ partitions, the Jacobian matrix is block diagonal in the partitions $J(\mathbf{X}^{(1)},\mathbf{X}^{(2)}) = J(\mathbf{X}^{(1)})\oplus J(\mathbf{X}^{(2)})$, which we use to define the split invariant coordinate transformations in the \textit{bipartite} (or \textit{mutual}) \textit{information} case.}
\begin{equation}
\gamma(\mathbf{X}^{(1)},\dots,\mathbf{X}^{(n)}) = \prod_{k=1}^n\gamma(\mathbf{X}^{(k)}).
\end{equation}
We then derive what we call the $n$-partite information (NPI),
\begin{equation}
I[\rho,\mathbf{X}^{(1)};\dots;\mathbf{X}^{(n)}] = \int dx\, p(x)\log\frac{p(x)}{\prod_{k=1}^np(x^{(k)})}.\label{npi}
\end{equation}
\indent  The combinatorial number of possible partitions of the spaces for $n\leq N$ splits is given by the combinatorics of Stirling numbers of the second kind \cite{Stirling}.  A Stirling number of the second kind $S(N,n)$ (often denoted as $\left\{\begin{matrix} N \\ n \end{matrix} \right\}$) gives the number of $n$ subsets one can form from a set of $N$ elements.  The definition in terms of binomial coefficients is given by,
\begin{equation}
\left\{\begin{matrix} N \\ n \end{matrix}\right\} = \frac{1}{n!}\sum_{i=0}^{n-1}(-1)^i\left(\begin{matrix}
n \\ i 
\end{matrix}\right)(n - i)^N.\label{stirling}
\end{equation}
Thus, the number of unique $n$-partite informations one can form from a set of $N$ variables is equal to $\left\{\begin{matrix}
N \\ n
\end{matrix}\right\}$.

Using (\ref{cond}) from the appendix, for any $n$-partite information $I[\rho,\mathbf{X}^{(1)};\dots;\mathbf{X}^{(k)};\dots;\mathbf{X}^{(n)}]$, where $n > 2$, we have the chain rule,
\begin{equation}
I[\rho,\mathbf{X}^{(1)};\dots;\mathbf{X}^{(k)};\dots;\mathbf{X}^{(n)}] = \sum_{k=2}^nI[\rho,\mathbf{X}^{(1)}\times\dots\times\mathbf{X}^{(k-1)};\mathbf{X}^{(k)}],\label{chain}
\end{equation}
where $I[\rho,\mathbf{X}^{(1)}\times\dots\times\mathbf{X}^{(k-1)};\mathbf{X}^{(k)}]$ is the mutual information between the subspace $\mathbf{X}^{(1)}\times\dots\times\mathbf{X}^{(k-1)}$ and the subspace $\mathbf{X}^{(k)}$.

\subsection{Remarks on the upper-bound of TC}
The TC provides an upper-bound for any choice of $n$-partition information, i.e. any $n$-partite information in which $n < N$ necessarily satisfies,
\begin{equation}
I[\rho,\mathbf{X}^{(1)};\dots;\mathbf{X}^{(n)}] \leq I[\rho,\mathbf{X}].
\end{equation}  
This can be shown by using the decomposition of the TC into continuous Shannon entropies which was discussed in \cite{Watanabe},
\begin{equation}
I[\rho,\mathbf{X}] = \sum_{i=1}^NS[\rho_{i},\mathbf{X}_i] - S[\rho,\mathbf{X}],\label{decomp}
\end{equation}
where the continuous Shannon entropy\footnote{While it is true that the continuous Shannon entropy is not coordinate invariant, the particular combinations used in this paper are, due to the TC and $n$-partite information being relative entropies themselves.} $S[\rho,\mathbf{X}]$ is,
\begin{equation}
S[\rho,\mathbf{X}] = -\int dx\, p(x)\log p(x).\label{contshannon}
\end{equation}
Likewise for any $n$-partition we have the decomposition,
\begin{equation}
I[\rho,\mathbf{X}^{(1)};\dots;\mathbf{X}^{(n)}] = \sum_{k=1}^nS[\rho^{{(k)}},\mathbf{X}^{(k)}] - S[\rho,\mathbf{X}].\label{kthdecomp}
\end{equation}
Since we in general have the inequality \cite{CoverThomas_Book} for entropy,
\begin{equation}
S[\rho_{ij},\mathbf{X}_i\times\mathbf{X}_j] \leq S[\rho_i,\mathbf{X}_i] + S[\rho_{j},\mathbf{X}_j],\label{inequal}
\end{equation}
then we also have that for any $k$-th partition of a set of $N$ variables, that the $N_k$ exhaustive internal partitions (i.e. $\sum_{k=1}^n N_k=N$) of $\mathbf{X}^{(k)}=\mathbf{X}^{(k^{(1)})}\times...\times\mathbf{X}^{(k^{(N_k)})}$ satisfy,
\begin{equation}
S[\rho^{(k)},\mathbf{X}^{(k)}] \leq \sum_{i=1}^{N_k}S[\rho^{(k^{(i)})},\mathbf{X}^{(k^{(i)})}].\label{kthinequal}
\end{equation}
Using (\ref{kthinequal}) in (\ref{kthdecomp}), we then have that for any $n$-partite information,
\begin{align}
I[\rho,\mathbf{X}^{(1)};\dots;\mathbf{X}^{(n)}] &= \sum_{k=1}^nS[\rho^{(k)},\mathbf{X}^{(k)}] - S[\rho,\mathbf{X}]\nonumber\\
&\leq \sum_{k=1}^n\left(\sum_{i=1}^{N_k}S[\rho^{(k^{(i)})},\mathbf{X}^{(k^{(i)})}]\right) - S[\rho,\mathbf{X}] = I[\rho,\mathbf{X}].\label{TCineq}
\end{align}
Thus, the Total Correlations are always greater than or equal to any correlations between any $n$-partite splitting of $\mathbf{X}$  Upper bounds for the discrete case was discussed in \cite{Merkh}.

\subsection{The bipartite (mutual) information}\label{mi}  
Perhaps the most studied special case of the NPI is the mutual information (MI), which is the smallest possible $n$-partition one can form.  As was discussed in the introduction, it is useful in inference tasks and was the first quantity to really be defined and exploited \cite{Shannon_Book} out of the general class of $n$-partite informations.\\
\indent  To analyze the mutual information, consider first relabeling the total space as $\mathbf{Z} = \mathbf{X}_1\times\dots\times\mathbf{X}_N$ to match the common notation in MI literature.  The \textit{bipartite information} considers only two subspaces, $\mathbf{X}\subset \mathbf{Z}$ and $\mathbf{Y}\subset\mathbf{Z}$, rather than all of them.  These two subspaces define a bipartite split in the proposition space such that $\mathbf{X}\cap\mathbf{Y} = \emptyset$ and $\mathbf{X}\times\mathbf{Y} = \mathbf{Z}$.  This results in turning the product marginal into,
\begin{equation}
m(x) \xrightarrow{MI} m(x,y) = p(x)p(y),\label{bipartite}
\end{equation}
where $x \in \mathbf{X}$ and $y \in \mathbf{Y}$.  Finally, we arrive at the functional that we will label by its split as,
\begin{equation}
I[\rho,\mathbf{X};\mathbf{Y}] = \int dxdy\, p(x,y)\log\frac{p(x,y)}{p(x)p(y)},
\end{equation}
which is the \textit{mutual information}.  Since the marginal space is split into two distinct subspaces, the mutual information only quantifies the correlations \textit{between} the two subspaces and not between all the variables as is the case with the \textit{total correlation} for a given split.  Whenever the total space $\mathbf{Z}$ is two-dimensional, the total correlation and the mutual information coincide.\\
\indent  One can derive the mutual information by using the same steps as in the total correlation derivation above, except replacing the independence condition in (\ref{min}) with the bipartite marginal in (\ref{bipartite}).  The goal is the same as (\ref{derive}) except that the MI \textit{ranks the distributions $\rho$ according to the correlations between two subspaces of propositions}, rather than within the entire proposition space.

\subsection{The discrete total correlation}
One may derive a discrete total correlation and discrete NPI by starting from equation (\ref{36}),
\begin{equation}
I[\rho,\mathcal{X}] = \sum_{i=1}^{|\mathcal{X}|}F_i(P(x_i)),
\end{equation}
and then following the same arguments without taking the continuous limit after DC1 was imposed. 

The inferential transformations explored in \hyperref[class]{Section II} are somewhat different for discrete distributions.  Coordinate transformations are replaced by general \textit{reparameterizations},\footnote{An example of such a discrete reparameterization (or discrete coordinate transformation) is intuitively used in coin flipping experiments -- the outcome of coin flips may be parameterized with (-1,1) or equally with (0,1) to represent tails verses heads outcomes, respectively.} in which one defines a bijection between sets,
\begin{align}
f:\mathcal{X} &\rightarrow \mathcal{X}'\nonumber\\
x_i &\mapsto f(x_i) = x'_i.\label{reparameterization}
\end{align}  
Like with general coordinate transformations (\ref{i}), if $f(x_i)$ is a bijection then we equate the probabilities,
\begin{equation}
P(x_i) = P(f(x_i)) = P(x'_i).\label{d_i}
\end{equation}
Since $x_i$ is simply a label for a proposition, we enforce that the probabilities associated to it are independent of the choice of label. One can define discrete split coordinate invariant transformations analogous to the continuous case. Using discrete split invariant coordinate transformations, the index $i$ is removed from $F_i$ above, which is analogous to the removal of the $x$ coordinate dependence in the continuous case. The functional equation for the $\log$ is found and solved analogously by imposing DC2. The discrete TC is then found and the discrete NPI may be argued.

The other transformations in \hyperref[class]{Section II} remain the same, except for replacing integrals by sums. In the above subsections, the continuous relative entropy is replaced by the discrete version,
\begin{equation}
S[P,Q] = -\sum_{x_i}P(x_i)\log\frac{P(x_i)}{Q(x_i)}.
\end{equation}
The discrete MI is extremely useful for dealing with problems in communication theory, such as noisy-channel communication and Rate-Distortion theory \cite{CoverThomas_Book}.  It is also reasonable to consider situations where one has combinations of discrete and continuous variables.  One example is the binary category case \cite{Carrara_Ernst}.

\section{Sufficiency}\label{sufficiency}
There is a large literature on the topic of sufficiency \cite{CoverThomas_Book,Kullback,Ay} which dates back to work originally done by Fisher \cite{Fisher_1}.  Some have argued that the idea dates back to even Laplace \cite{Stigler}, a hundred years before Fisher.  What both were trying to do ultimately was determine whether one could find simpler statistics that contain all of the required information to make equal inferences about some parameter.  

Let $p(x,\theta) = p(x)p(\theta|x) = p(\theta)p(x|\theta)$ be a joint distribution over some variables $\mathbf{X}$ and some parameters we wish to infer $\mathbf{\Theta}$.  Consider then a function $y = f(x)$, and also the joint density,
\begin{equation}
p(x,y,\theta) = p(x,y)p(\theta|x,y) = p(x)p(\theta|x,y)\delta(y - f(x)).
\end{equation}
If $y$ is a sufficient statistic for $x$ with respect to $\theta$, then the above equation becomes,
\begin{equation}
p(x,y,\theta) = p(x)p(\theta|y)\delta(y - f(x))\label{sufficient},
\end{equation}
and the conditional probability $p(\theta|x,y) = p(\theta|y)$ doesn't depend on $x$ because $y$ is sufficient.  Fisher's factorization theorem states that a sufficient statistic for $\theta$ will give the following relation,
\begin{equation}
p(x|\theta) = f(y|\theta)g(x),
\end{equation}
where $f$ and $g$ are functions that are not necessarily probabilities; i.e. they are not normalized with respect to their arguments, however since the left hand side is certainly normalized with respect to $x$, then the right hand side must be as well.  To see this, we can rewrite  eq. (\ref{sufficient}) in terms of the distributions,
\begin{equation}
p(x,y,\theta) = p(x,\theta)p(y|x,\theta) = p(\theta)p(x|\theta)\delta(y - f(x))\label{sufficient_2}.
\end{equation}
Equating eqs. (\ref{sufficient}) and (\ref{sufficient_2}) we find,
\begin{equation}
p(x)p(\theta|y) = p(\theta)p(x|\theta),
\end{equation}
which gives the general result,
\begin{equation}
p(x|\theta) = \frac{p(\theta|y)}{p(\theta)}p(x) = \frac{p(y|\theta)}{p(y)}p(x).
\end{equation}
We can then identify $g(x) = p(x)$ which only depends on $x$ and $f(y|\theta) = p(y|\theta)/p(y)$ which is the ratio of two probabilities and hence, not normalized with respect to $y$.

\subsection{A new definition of sufficiency}
While the notion of a sufficient statistic is useful, how can we quantify the sufficiency of a statistic which is not completely sufficient but only partially?  What about for the case of generic $n$-partite systems?  The $n$-partite information can provide an answer.  We first begin with the bi-partite case.  

Our design derivation shows that the MI is uniquely designed for quantifying the global correlations in the bi-partite case. Because the correlations between two variables indicate how informative one variable is toward the inference of the other, a change in the MI indicates a change in ones ability to make such inferences over the entire spaces of both variables. Thus, we can use this interpretation toward quantifying statistical sufficiency in terms of the change in the amount of correlations in a global sense.

Consider an arbitrary continuous function $f:\mathbf{X}\rightarrow\mathbf{X}'$, which we call a \textit{statistic} of $\mathbf{X}$.  We define the \textit{sufficiency} of the statistic $f(\mathbf{X})$ with respect to another space $\Theta$ for a bi-partite system as simply the ratio of mutual informations,
\begin{equation}
\mathrm{suff}_{\Theta}[f(\mathbf{X})] \stackrel{\mathrm{def}}{=} \frac{I[\rho',f(\mathbf{X});\Theta]}{I[\rho,\mathbf{X};\Theta]}\label{suff},
\end{equation}
which is always bounded by $0 \leq \mathrm{suff}_{\mathbf{Y}}(f) \leq 1$ due to the data processing inequality in \hyperref[data]{Appendix C.3} and $\rho'$ is the distribution defined over the joint space $f(\mathbf{X})\times\Theta$. In this problem space, it is assumed that there exists correlation between $(\mathbf{X},\Theta)$, i.e. $I[\rho,\mathbf{X};\Theta]>0$, at least before the statistic is checked for sufficiency. Statistics for which $\mathrm{suff}_{\Theta}[f(\mathbf{X})] = 1$ are called \textit{sufficient} and correspond to the definition given by Fisher.  We can see this by appealing to the special case $p(x,f(x),\theta) = p(x)p(\theta|x)\delta(y - f(x))$ for some statistic $y = f(x)$.  It is true that,
\begin{equation}
p(y) = \int dx\, p(x,y) = \int dx\, p(x)\delta(y - f(x)),
\end{equation}  
so that when $p(\theta|y) = p(\theta|x)$,
\begin{align}
I[\rho',f(\mathbf{X});\Theta] &= \int dyd\theta\, p(y)p(\theta|y)\log\frac{p(\theta|y)}{p(\theta)}\nonumber\\
&= \int dxdyd\theta p(x)\delta(y - f(x))p(\theta|x)\log\frac{p(\theta|x)}{p(\theta)}\nonumber\\
&= I[\rho,\mathbf{X};\Theta],
\end{align}
which is the criteria for $y$ to be a sufficient statistic.  With this definition of sufficiency (\ref{suff}) we have a way of evaluating maps $f(\mathbf{X})$ which attempt to preserve correlations between $\mathbf{X}$ and $\mathbf{Y}$.  These procedures are ubiquitous in machine learning \cite{Carrara_Ernst}, manifold learning and other inference tasks, although the explicit connection with statistical sufficiency has yet to be realized.  

\subsection{The $n$-partite sufficiency}
Like mutual information, the $n$-partite information can provide insights into sufficiency.  Let us first begin by stating a theorem.
\begin{theorem2}[$n$-partite information inequality]\label{theorem1}
	Let $\mathbf{X} = \mathbf{X}_1\times\dots\times\mathbf{X}_N$ be a collection of $N$ subspaces and let $\{\mathbf{X}^{(k)}\}_n$ be an $n$-partite system of $\mathbf{X}$.  Then, for any function $f:\mathbf{X}^{(k)}\rightarrow\mathbf{X}^{(k)'}$ which acts on one of the $n$-partite subspaces, we have the following inequality,
	\begin{equation}
	I[\rho,\mathbf{X}^{(1)};\dots;\mathbf{X}^{(k)};\dots;\mathbf{X}^{(n)}] \geq I[\rho',\mathbf{X}^{(1)};\dots;f(\mathbf{X}^{(k)});\dots;\mathbf{X}^{(n)}].\label{npartineq}
	\end{equation}
\end{theorem2}
The proof of the above theorem can be written in analogy to the data processing inequality derivation in \hyperref[data]{Appendix C.3}.
\begin{proof}
	Let $\{\mathbf{X}^{(k)}\}_n$ be an $n$-partite system for a collection of $N$ variables, $\mathbf{X} = \mathbf{X}_1\times\dots\times\mathbf{X}_N$.  The joint distribution can be written,
	\begin{equation}
	p(x_1,\dots,x_N) = p(x_1)p(x_2|x_1)\cdots p(x_N|x_1,\dots,x_{N-1}),
	\end{equation}
	and the $n$-partite marginal $m(x)$ is,
	\begin{equation}
	m(x) = \prod_{k=1}^np(x^{(k)}),
	\end{equation}
	where $x^{(k)} \in \mathbf{X}^{(k)}$ is the $k$th collection in $\{\mathbf{X}^{(k)}\}_n$.  Thus the $n$-partite information is written,
	\begin{equation}
	I[\rho,\mathbf{X}^{(1)};\dots;\mathbf{X}^{(n)}] = \int dx\, p(x)\log\frac{p(x)}{\prod_{k=1}^np(x^{(k)})}.
	\end{equation}
	Consider now that we define the function,
	\begin{align}
	f_{x^{(k)'}}:\mathbf{X}^{(k)} &\rightarrow \mathbf{X}^{(k)'}\nonumber\\
	x^{(k)} &\mapsto f_{x^{(k)'}}(x^{(k)}) = x^{(k)'},\label{suff2}
	\end{align}
	which takes the collection of variables $\mathbf{X}^{(k)}$ to some other collection $\mathbf{X}^{(k)'}$.  Thus, we have the joint distribution,
	\begin{equation}
	p(x^{(k)},x^{(k)'}) = p(x^{(k)})p(x^{(k)'}|x^{(k)}) = p(x^{(k)})\delta(x^{(k)'} - f_{x^{(k)'}}(x^{(k)})).
	\end{equation}
	Now, consider the case in which the variables $\mathbf{X}^{(k)'}$ are combined with the original variables $\mathbf{X}^{(k)}$ in their respective partition as a Cartesian product, $\mathbf{X}^{(k)} \rightarrow \mathbf{X}^{(k)}\times\mathbf{X}^{(k)'}$, so that the $n$-partite information becomes,
	\begin{align}
	I[\rho,\mathbf{X}^{(1)};\dots;\mathbf{X}^{(k)}\times\mathbf{X}^{(k)'};\dots;\mathbf{X}^{(n)}] &= I[\rho',\mathbf{X}^{(1)};\dots;\mathbf{X}^{(k)};\dots;\mathbf{X}^{(n)}]\nonumber\\
	&+ \tilde{I}[\rho'',\mathbf{X}^{(1)};\dots;\mathbf{X}^{(k)'};\dots;\mathbf{X}^{(n)}|\mathbf{X}^{(k)}],\label{npart1}
	\end{align}
	where the second term on the right hand side is,
	\begin{align}
	\tilde{I}[\rho'',\mathbf{X}^{(1)};\dots;\mathbf{X}^{(n)}|\mathbf{X}^{(k)}] &= \int dx\, p(x)\log\frac{p(x^{(k)'}|x^{(1)},\dots,x^{(k)},\dots,x^{(n)})}{p(x^{(k)'}|x^{(k)})}\nonumber\\
	&= 0,
	\end{align}
	since $p(x^{(k)'}|x^{(1)},\dots,x^{(k)},\dots,x^{(n)}) = p(x^{(k)'}|x^{(k)}) = \delta(x^{(k)'} - f_{x^{(k)'}}(x^{(k)}))$.  Consider however that we break up the $n$-partite information in (\ref{npart1}) by first removing $\mathbf{X}^{(k)}$ instead of $\mathbf{X}^{(k)'}$,
	\begin{align}
	I[\rho,\mathbf{X}^{(1)};\dots;\mathbf{X}^{(k)}\times\mathbf{X}^{(k)'};\dots;\mathbf{X}^{(n)}] &= I[\rho',\mathbf{X}^{(1)};\dots;\mathbf{X}^{(k)'};\dots;\mathbf{X}^{(n)}]\nonumber\\
	&+ \tilde{I}[\rho'',\mathbf{X}^{(1)};\dots;\mathbf{X}^{(k)};\dots;\mathbf{X}^{(n)}|\mathbf{X}^{(k)'}],\label{xpart1}
	\end{align}
	where again the second term is,
	\begin{align}
	\tilde{I}[\rho'',\mathbf{X}^{(1)};\dots;\mathbf{X}^{(n)}|\mathbf{X}^{(k)'}] &= \int dx\, p(x)\log\frac{p(x^{(k)}|x^{(1)},\dots,x^{(k)'},\dots,x^{(n)})}{p(x^{(k)}|x^{(k)'})}\nonumber\\
	&= I[\rho'',\mathbf{X}^{(k)};\tilde{\mathbf{X}}|\mathbf{X}^{(k)'}] \geq 0,
	\end{align}
	and where $\tilde{\mathbf{X}} = \mathbf{X}\backslash\mathbf{X}^{(k)}$.  Thus we have that,
	\begin{align}
	I[\rho,\mathbf{X}^{(1)};\dots;\mathbf{X}^{(k)}\times\mathbf{X}^{(k)'};\dots;\mathbf{X}^{(n)}] &= I[\rho',\mathbf{X}^{(1)};\dots;\mathbf{X}^{(k)};\dots;\mathbf{X}^{(n)}]\nonumber\\
	&= I[\rho'',\mathbf{X}^{(1)};\dots;\mathbf{X}^{(k)'};\dots;\mathbf{X}^{(n)}]\nonumber\\
	&+ I[\rho''',\mathbf{X}^{(k)};\tilde{\mathbf{X}}|\mathbf{X}^{(k)'}],
	\end{align}
	and thus we have that for a general function $f_{x^{(k)'}}$ of the $k$th-partition, the inequality,
	\begin{equation}
	I[\rho,\mathbf{X}^{(1)};\dots;\mathbf{X}^{(k)};\dots;\mathbf{X}^{(n)}] \geq I[\rho',\mathbf{X}^{(1)};\dots;f(\mathbf{X}^{(k)});\dots;\mathbf{X}^{(n)}],
	\end{equation}
	which proves the \hyperref[theorem1]{theorem}.
\end{proof}
The above theorem provides a ranking of transformation functions $(f_{x^{(1)'}},f_{x^{(2)'}},\dots,f_{x^{(n)'}})$ of the $n$-partitions,
\begin{align}
I[\rho,\mathbf{X}^{(1)};\dots;\mathbf{X}^{(n)}] &\geq I[\rho',f(\mathbf{X}^{(1)});\dots;\mathbf{X}^{(n)}]\nonumber\\
&\geq I[\rho'',f(\mathbf{X}^{(1)});f(\mathbf{X}^{(2)});\dots;\mathbf{X}^{(n)}]\nonumber\\
&\geq \quad\dots\nonumber\\
&\geq I[\rho^{'\cdots '},f(\mathbf{X}^{(1)});f(\mathbf{X}^{(2)});\dots;f(\mathbf{X}^{(n)})].\label{suff5}
\end{align}
The action of any set of functions $\{f\}$ can only ever decrease the amount of $n$-partite correlations, much as in the data processing inequality.  We define the sufficiency of a set of functions $\{f\}$ analogously to (\ref{suff}).  Consider that we generate a set of functions for $m$ of the partitions, leaving $n-m$ partitions alone.  Then the sufficiency of the set of functions $\{f\}$ which act on the subspace $\{\mathbf{X}^{(k)}\}_m$ with respect to the remaining $m-n$ partitions is given by,
\begin{equation}
\mathrm{suff}_{\tilde{\mathbf{X}}}(\{f\}) \stackrel{\mathrm{def}}{=} \frac{I[\rho',f(\mathbf{X}^{(1)});\dots;f(\mathbf{X}^{(m)});\mathbf{X}^{(m+1)};\dots;\mathbf{X}^{(n)}]}{I[\rho,\mathbf{X}^{(1)};\dots;\mathbf{X}^{(n)}]},
\end{equation}
where $\tilde{\mathbf{X}} = \{\mathbf{X}^{(k)}\}_n\backslash\{\mathbf{X}^{(k)}\}_m$.  Like the bi-partite sufficiency, the $n$-partite sufficiency is bounded between zero and one due to the $n$-partite information inequality.

\subsection{The n-partite joint processing inequality}%The multi-partite inequality

Using the results from eqs. (\ref{TCineq}) and (\ref{npartineq}), we can express a general result which we call the $n$-partite joint processing inequality.  While the $n$-partite inequality concerns functions which act individually within the $n$-partite spaces, we can generalize this notion to functions which act over the entire variable space, i.e. functions which jointly process partitions.

\begin{theorem2}[$n$-partite joint processing inequality]\label{theorem3}
	Let $\mathbf{X} = \mathbf{X}_1\times\dots\times\mathbf{X}_N$ be a collection of $N$ subspaces and let $\{\mathbf{X}^{(k)}\}_n$ be an $n$-partite system of $\mathbf{X}$.  Then, for any function $f:\mathbf{X}^{(k)}\times\mathbf{X}^{(\ell)}\rightarrow\mathbf{X}^{(k)'}$ which acts on two of the $n$-partite subspaces, we have the following inequality,
	\begin{equation}
	I[\rho,\mathbf{X}^{(1)};\dots;\mathbf{X}^{(k)};\mathbf{X}^{(\ell)};\dots;\mathbf{X}^{(n)}] \geq I[\rho',\mathbf{X}^{(1)};\dots;f(\mathbf{X}^{(k)},\mathbf{X}^{(\ell)});\dots;\mathbf{X}^{(n)}].\label{njointeq}
	\end{equation}
\end{theorem2}

\begin{proof}
	Consider the case of three partitions, $\mathbf{X} = \mathbf{X}^{(1)}\times\mathbf{X}^{(2)}\times\mathbf{X}^{(3)}$.  Then, using a result from \hyperref[margagain]{appendix C.1.3}, eq. (\ref{cond}), it is true that,
	\begin{equation}
	I[\rho,\mathbf{X}^{(1)};\mathbf{X}^{(2)};\mathbf{X}^{(3)}] \geq I[\rho',\mathbf{X}^{(1)}\times\mathbf{X}^{(2)};\mathbf{X}^{(3)}].
	\end{equation}
	Thus, together with the $n$-partite information inequality (\ref{npartineq}), any function $f_{x^{(1,2)'}}$ which combines two partitions necessarily satisfies,
	\begin{equation}
	I[\rho,\mathbf{X}^{(1)};\mathbf{X}^{(2)};\mathbf{X}^{(3)}] \geq I[\rho',f_{x^{(1,2)'}}\left(\mathbf{X}^{(1)},\mathbf{X}^{(2)}\right);\mathbf{X}^{(3)}].\label{suff4}
	\end{equation}
\end{proof}
Thus, an analogous continuous definition of the sufficiency follows for $n$-partitions of $m$ joint statistic functions.  Consider a function $f_{x^{(1,\dots,m)}}$ which combines the first $m$ partitions of an $n$-partite system,
\begin{align}
f_{x^{(1,\dots,m)}}:\mathbf{X}^{(1)}\times\dots\times\mathbf{X}^{(m)} &\rightarrow \mathbf{X}'\nonumber\\
(x^{(1)},\dots,x^{(m)}) &\mapsto f_{x^{(1,\dots,m)}}(x^{(1)},\dots,x^{(m)}) = x'.\label{suff3}
\end{align}
Then we define the sufficiency of the map $f_{x^{(1,\dots,m)}}$ with respect to the remaining $m-n$ partitions as the ratio,
\begin{equation}
\mathrm{suff}_{\tilde{\mathbf{X}}}(f_{x^{(1,\dots,m)}}) \stackrel{\mathrm{def}}{=}\frac{I[\rho',\mathbf{X}';\mathbf{X}^{(m+1)};\dots;\mathbf{X}^{(n)}]}{I[\rho,\mathbf{X}^{(1)};\dots;\mathbf{X}^{(m)};\mathbf{X}^{(m+1)};\dots;\mathbf{X}^{(n)}]}.
\end{equation}
The are many possible combinations of maps of the form (\ref{suff3}) and (\ref{suff2}) that may analogously be expressed with a continuous notion of sufficiency; however, for the application of successive functions, one will always find a nested set of inequalities of the form (\ref{suff4}) and (\ref{suff5}).

\subsection{The likelihood ratio}Here we will associate the invariance of MI to invariance of type I and type II errors.  Consider a binary decision problem in which we have some set of discriminating variables $\mathbf{X}$, following a mixture of two distributions (e.g. signal and background) labeled by a parameter $\theta = \{s,b\}$.  The inference problem can then be cast in terms of the joint distribution $p(x,\theta) = p(x)p(\theta|x)$.  According to the Neyman-Pearson lemma \cite{Neyman-Pearson}, the likelihood ratio,
\begin{equation}
\Phi(x) = \frac{\mathcal{L}(s|x)}{\mathcal{L}(b|x)} = \frac{p(x|s)}{p(x|b)}\label{likelihood},
\end{equation}
gives a sufficient statistic for the significance level,
\begin{equation}
\alpha(x) = P(\Phi(x) \leq \eta|b)\label{sig},
\end{equation}
where $b = H_0$ is typically associated to the \textit{null hypothesis}.  This means that the likelihood ratio (\ref{likelihood}) will allow us to determine if the data $\mathbf{X}$ satisfies the significance level in (\ref{sig}).  Given Bayes' theorem, the likelihood ratio is equivalent to,
\begin{equation}
\Phi(x) = \frac{p(x|s)}{p(x|b)} = \frac{p(b)}{p(s)}\frac{p(s|x)}{p(b|x)} = \frac{p(b)}{p(s)}\Pi(x),
\end{equation}
which is the \textit{posterior} ratio and is just as good of a statistic as the likelihood ratio, since $p(b)/p(s)$ is a constant for all $x \in \mathbf{X}$.  If we then construct a sufficient statistic $y = f(x)$ for $\mathbf{X}$, such that,
\begin{equation}
\mathbf{Y} = f(\mathbf{X}) \quad \rightarrow \quad I[\rho',f(\mathbf{X});\theta] = I[\rho,\mathbf{X};\theta],
\end{equation}  
then the posterior ratios, and hence the likelihood ratios, are equivalent,
\begin{equation}
\Pi(f(x)) = \frac{p(s|f(x))}{p(b|f(x))} = \frac{p(s|x)}{p(b|x)} = \Pi(x),\label{sufficient2}
\end{equation}
and hence the significance levels are also invariant,
\begin{equation}
\alpha(f(x)) = P(\Phi(f(x)) \leq \eta|b) = P(\Phi(x) \leq \eta|b) = \alpha(x),
\end{equation}
and therefore the type I and type II errors will also be invariant.  Thus we can use the MI as a tool for finding the type I and type II errors for some unknown probability distribution by first constructing a sufficient statistic using some technique (typically a ML technique), and then finding the type I and type II errors on the simpler distribution.  Then, due to (\ref{sufficient2}), the errors associated to the simpler distribution will be equivalent to the errors of the original unknown one.\\
\indent  Apart from its invariance, we can also show another consequence of MI under arbitrary transformations $f(\mathbf{X})$ for binary decision problems.  Imagine that we successfully construct a sufficient statistic for $\mathbf{X}$.  Then, it is a fact that the likelihood ratios $\Phi(x)$ and $\Phi(f(x))$ will be equivalent for all $x \in \mathbf{X}$.  Consider that we adjust the probability of one value of $p(\theta|f(x_i))$ by shifting the relative weight of signal and background for that particular value $f(x_i)$,
\begin{align}
p(s|f(x_i)) \rightarrow p'(s|f(x_i)) &= p(s|f(x_i)) + \delta p\nonumber\\
p(b|f(x_i)) \rightarrow p'(b|f(x_i)) &= p(b|f(x_i)) - \delta p,
\end{align}    
where $\delta p$ is some small change, so that the particular value of
\begin{equation}
\Pi'(f(x_i)) = \frac{p'(s|f(x_i))}{p'(b|f(x_i))} = \frac{p(s|f(x_i))+\delta p}{p(b|f(x_i)) -\delta p} \neq \Pi(f(x_i)),
\end{equation}
which is not equal to the value given from the sufficient statistic.  Whether the value $\Pi'(f(x_i))$ is larger or smaller than $\Pi(f(x_i))$, in either case either the number of type I or type II errors will increase for the distribution with $p'(\theta|f(x_i))$ replaced for the sufficient value $p(\theta|f(x_i))$.  Therefore, for any distribution given by the joint space $\mathbf{X}\times\Theta$, the MI determines the type I and type II error for any statistic on the data $\mathbf{X}$.

\section{Discussion of alternative measures of correlation}
The design derivation in this paper puts the various NPI functionals on an equal foundational footing as the relative entropy.  This begs the question as to whether other similar information theoretic quantities can be designed along similar lines.  Some quantities that come to mind are \textit{$\alpha$-mutual information} \cite{Verdu}, \textit{multivariate-mutual information} \cite{McGill,Hu}, \textit{directed information} \cite{Massey}, \textit{transfer entropy} \cite{Schreiber} and \textit{causation entropy} \cite{Sun1,Sun2,Cafaro}.  

The $\alpha$-mutual information \cite{Verdu} belongs to the family of functionals that fall under the name \textit{R\'{e}nyi entropies} \cite{Renyi} and their close cousin \textit{Tsallis entropy} \cite{Tsallis1,Tsallis2}.  Tsallis proposed his entropy functional as an attempted generalization for applications of statistical mechanics, however the probability distributions that it produces can be generated from the standard MaxEnt procedure \cite{Caticha_Book} and do not require a new thermodynamics\footnote{For some discussion on this topic see \cite{Caticha_Book} page 114.}.  Likewise, R\'{e}nyi's family of entropies attempts to generalize the relative entropy for generic inference tasks, which inadvertently relaxes some of the design criteria concerning independence.  Essentially, R\'{e}nyi introduces a set of parameterized entropies $S_{\eta}[p,q]$, with parameter $\eta$, which leads to the weakening of the independent subsystem additivity criteria.  Imposing that these functionals then obey subsystem independence immediately constrains $\eta = 0$ or $\eta = -1$, and reduces them back to the standard relative entropy, i.e. $S_{\eta=0}[p,q] \equiv S[p,q]$ and $S_{\eta=-1}[p,q] \equiv S[q,p]$.  Without a strict understanding of what it means for subsystems to be independent, one cannot conduct reasonable science.  Thus, such ``generalized'' measures of correlation (such as the $\alpha$-mutual information \cite{Verdu}) which abandon subsystem independence cannot be trusted.

Defining multivariate-mutual information (MMI) is an attempt to generalize the standard MI to a case where several sets of propositions are compared with each other, however this is different than the total correlation which we designed in this paper.  For example, given three sets of propositions $\mathbf{X},\mathbf{Y}$ and $\mathbf{Z}$, the \textit{multivariate-mutual information} (not to be confused with the \textit{multi-information} \cite{Studeny} which is another name for total correlation) is,
\begin{equation}
MMI[\rho,\mathbf{X};\mathbf{Y};\mathbf{Z}] \stackrel{\mathrm{def}}{=} I[\rho,\mathbf{X};\mathbf{Y}] - I[\rho,\mathbf{X};\mathbf{Y}|\mathbf{Z}]
\end{equation}
One difficulty with this expression is that it can be negative, as was shown by Hu \cite{Hu}.  Thus, defining a minimum MMI is not possible in general, which suggests that a design derivation of MMI requires a different interpretation.  Despite these difficulties, there have been several recent successes in the study of MMI including Bell \cite{Bell} and Baudot et. al. \cite{Baudot12,Baudot22} who studied MMI in the context of algebraic topology.\\
\indent  Another extension of mutual information is transfer entropy, which was first introduced by Schreiber \cite{Schreiber} and is a special case of \textit{directed information} \cite{Massey}.  Transfer entropy is a conditional mutual information which attempts to quantify the amount of ``information'' that flows between two time-dependent random processes.  Given a set of propositions which are dynamical, such that $\mathbf{X} = \mathbf{X}(t)$ and $\mathbf{Y} = \mathbf{Y}(t)$, so that at time $t_i$ the propositions take the form $\mathbf{X}(t_i) \stackrel{\mathrm{def}}{=}\mathbf{X}_{t_i}$ and $\mathbf{Y}(t_i) \stackrel{\mathrm{def}}{=}\mathbf{Y}_{t_i}$, then the \textit{transfer entropy} (TE) between $\mathbf{X}(t)$ and $\mathbf{Y}(t)$ at time $t_i$ is defined as the conditional mutual information,
\begin{equation}
T_{\mathbf{X}\rightarrow\mathbf{Y}} \stackrel{\mathrm{def}}{=} I[\rho,\mathbf{Y}_{t_i};\mathbf{X}_{t_{j < i}}|\mathbf{Y}_{t_{j<i}}].
\end{equation}
The notation $t_{j<i}$ refers to all times $t_j$ before $t_i$.  Thus, the TE is meant to quantify the influence of a variable $\mathbf{X}$ on predicting the state $\mathbf{Y}_{t_i}$ when one already knows the history of $\mathbf{Y}$, i.e. it quantifies the amount of independent correlations provided by $\mathbf{X}$.  Given that TE is a conditional mutual information, it does not require a design derivation independent of the MI.  It can be justified on the basis of the discussion around (\ref{chain}).  Likewise the more general \textit{directed information} is also a conditional MI and hence can be justified in the same way.\\
\indent  Finally, the definition of \textit{causation entropy} \cite{Sun1,Sun2,Cafaro} can also be expressed as a conditional mutual information.  Causation entropy (CE) attempts to quantify time-dependent correlations between nodes in a connected graph and hence generalize the notion of transfer entropy.  Given a set of nodes $\mathbf{X},\mathbf{Y}$ and $\mathbf{Z}$, the causation entropy between two subsets conditioned on a third is given by,
\begin{equation}
C_{\mathbf{X}\rightarrow\mathbf{Y}|\mathbf{Z}} \stackrel{\mathrm{def}}{=} I[\rho,\mathbf{Y}_{t_i};\mathbf{X}_{t_{j < i}} | \mathbf{Z}_{t_{j<i}}].
\end{equation}
The above definition reduces to the transfer entropy whenever the set of variables $\mathbf{Z} = \mathbf{Y}$.  As was shown by Sun et. al. \cite{Sun3}, the causation entropy (CE) allows one to more appropriately quantify the causal relationships within connected graphs, unlike the transfer entropy which is somewhat limited.  Since CE is a conditional mutual information, it does not require an independent design derivation.  As with transfer entropy and directed information, the interpretation of CE can be justified on the basis of (\ref{chain}).

\section{Conclusions}
Using a design derivation, we showed that the TC is the functional designed to rank the global amount of correlations in a system between all of its variables.  We relied heavily on the PCC, which while quite simple, is restrictive enough to isolate the functional form of $I[\rho,\mathbf{X}]$ using eliminative induction. We enforced the PCC using two different methods as an additional measure of rigor (analytically through Taylor expanding and algebraically through the functional equation (\ref{F})).  The fact that both approaches lead to the same functional shows that the design criteria are highly constraining, and the total correlation is the unique solution. We generalized our solution to the $n$-partite information -- this global correlation quantifier can express the TC and MI as special cases.

Using our design derivation we were able to quantify the amount of global correlations and analyze the effect of inferential transformations in a new light.  Because the correlations between variables indicate how informative a set of variables are toward the inference of the others, a change in the global amount indicates a change in ones ability to make such inferences globally. Thus, we can use NPI to quantify statistical sufficiency in terms of the change in the amount of correlations over the entire joint variable spaces. This leads to a rigorous quantification of continuous statistical sufficiency that takes an upper bound of one when the Fisher sufficiency condition satisfied. %The continuous definition of sufficiency we present is simply the ratio of the NPI before and after a statistic is implemented.\\
% \indent     %

\paragraph{Acknowledgements}
We would like to thank A. Caticha for his invaluable support and guidance on this project.  We would also like to thank J. Ernst, S. Ipek and K. Knuth for many insightful and inspirational conversations.

\begin{appendix}
	\section{Coordinate Invariance}\label{coordinate}
	Consider a continuous space of propositions $\mathbf{X} \subseteq \mathbb{R}$ and an associated statistical manifold $\mathbf{\Delta}$ defined by (\ref{stat_man}).  A point in the statistical manifold is labeled $\rho \in \mathbf{\Delta}$ which is a map,
	\begin{align}
	\rho:\mathcal{P}(\mathbf{X})&\rightarrow [0,1]\nonumber\\
	x&\mapsto p(x).
	\end{align}
	Consider now the cumulative distribution of $\mathbf{X}$ for some value of $\tilde{x} \in \mathbf{X}$,
	\begin{equation}
	P(\mathbf{X} \leq \tilde{x}) = \int_{x\leq \tilde{x}}dx\, p(x).\label{cum}
	\end{equation}
	One can always recover the density $p(x)$ by differentiating (\ref{cum}) with respect to $x$,
	\begin{equation}
	p(x) = \frac{\partial}{\partial x}P(\mathbf{X} \leq \tilde{x}) = \frac{\partial}{\partial x}\int_{x\leq \tilde{x}}dx\, p(x).\label{density}
	\end{equation} 
	If we now consider a smooth bijection $f: \mathbf{X} \rightarrow \mathbf{Y}$, so that all the first derivatives exist, then it is true that $f(x)$ will always be an increasing or decreasing function of x.  It is true then that if $x \leq \tilde{x}$, then $f(x) \leq f(\tilde{x})$.  It will then be true that the cumulative distributions will be equal,
	\begin{equation}
	P(\mathbf{X}\leq\tilde{x}) = P(f(\mathbf{X})\leq f(\tilde{x})) = P(\mathbf{Y}\leq \tilde{y}).
	\end{equation}
	Therefore, the cumulative distribution distribution over $\mathbf{Y}$ up to some value $\tilde{y} \in \mathbf{Y}$ is,
	\begin{equation}
	P(\mathbf{Y} \leq \tilde{y}) = \int_{y\leq \tilde{y}}dy\, p'(y) = \int_{x\leq \tilde{x}}dx\, p(x),
	\end{equation}
	so that the density $p(x)$ from (\ref{density}) becomes,
	\begin{equation}
	p(x) = \frac{\partial}{\partial x}P(\mathbf{Y} \leq \tilde{y}) = \frac{\partial}{\partial x}\int_{y\leq \tilde{y}}dy\, p'(y) = \frac{\partial y}{\partial x}p'(y),\label{trans}
	\end{equation}
	which shows that the densities transform as,
	\begin{equation}
	p(x)dx = p'(y)dy.
	\end{equation}
	For a generic $n$-dimensional space of propositions $\mathbf{X} \subset \mathbb{R}^n$, then density (\ref{density}) becomes,
	\begin{equation}
	p(x) = \frac{\partial^n}{\partial x_1\dots\partial x_n}\int_{x\leq\tilde{x}}dx\, p(x) = \frac{\partial^n}{\partial x_1\dots\partial x_n}\int_{y\leq \tilde{y}}dy\, p'(y),
	\end{equation}
	where we suppressed the notation in the measure $d^nx = dx$ and $d^ny = dy$.  Then, the transformation (\ref{trans}) becomes,
	\begin{equation}
	p(x) = p'(y)\gamma(y)= p'(y)\left|\frac{\partial y}{\partial x}\right|,
	\end{equation}
	where $\left|\frac{\partial y}{\partial x}\right|$ is the Jacobian of $\mathbf{X}$ and $\mathbf{Y}$.
	\section{Watanabe's theorem}\label{watanabe}
	A theorem presented by Watanabe \cite{Watanabe} concerns the grouping property of entropy and its relation to total correlation.  The theorem was stated in \cite{Watanabe} on page 70.,\\
	\begin{theorem2}[Watanabe (1960)]
		The set of all variables in consideration is divided into subsets, and each subset is again subdivided into sub-subsets, et cetera, until finally the entire set is branched into individual variables.  Then, the sum of all correlations, each of which is defined with respect to a branching point, is independent of the way in which this branching procedure is made and is equal to the total correlation.
	\end{theorem2}
	Essentially, one can choose any set of $n$-partite splittings to a set of $N$ variables $\mathbf{X}$ and still arrive at the same total correlation by iterating over the splitting until all the variables have been individually split and adding the $n$-partite information's at each split. This can be proved easily using the grouping property of entropy.\\
	\begin{proof}
		To prove the theorem, let $\ell$ denote the level of the splitting sequence of the proposition space $\mathbf{X}$ so that $\mathbf{X}^{\ell}\subset \mathcal{P}(\mathbf{X})$ denotes the collection of all variables at the $\ell$th level.  Let $n_{\ell}$ denote the number of subsets in the $\ell$th split and let $\mathbf{X}_{k}^{\ell} \subset \mathbf{X}^{\ell}$ denote the $k$th subset of the subset $\mathbf{X}^{\ell}$.  Also let $n^k_{{\ell}}$ denote the number of subsets within the $k$th subset of the $\ell$th split so that $\sum_{k}n^k_{\ell} = n_{\ell}$.  We then have that,
		\begin{equation}
		\bigtimes_{k=1}^{n_{\ell}}\mathbf{X}^{\ell}_k = \mathbf{X}, \quad \mathrm{and} \quad \mathbf{X}^{\ell}_k\cap\mathbf{X}^{\ell}_j = \emptyset,\, \forall\, k \neq j.\label{intersect}
		\end{equation}
		Each subset $\mathbf{X}_{k}^{\ell}$ contains within it another subset of variables at the split level $(\ell + 1)$, so that we identify,
		\begin{equation}
		\mathbf{X}^{(\ell+1)}_k = \mathbf{X}^{\ell}_{k_i}, \quad \mathrm{where} \quad \bigcup_{i=1}^{n^k_{\ell}}\mathbf{X}_{k_i}^{\ell} = \mathbf{X}^{\ell}_k.\label{identify}
		\end{equation}  
		The combination of three indices $(\ell,k,i)$ defines a unique subset at level $(\ell+1,k)$.  We can write the $n$-partite information of the $k$th subset at level $\ell$ using (\ref{kthdecomp}) as,
		\begin{equation}
		I[\rho,\mathbf{X}^{\ell}_k] \stackrel{\mathrm{def}}{=} \sum_{i=1}^{n^k_{\ell}}S[\rho,\mathbf{X}_{k_i}^{\ell}] - S[\rho,\mathbf{X}^{\ell}_k].
		\end{equation}
		Continuing to the next level of the split $(\ell+1)$, we can also write down the $n$-partite information for each of the $n^k_{\ell}$ subspaces as,
		\begin{equation}
		I[\rho,\mathbf{X}^{\ell+1}_k] = \sum_{i=1}^{n^k_{\ell+1}}S[\rho,\mathbf{X}^{\ell+1}_{k_i}] - S[\rho,\mathbf{X}^{\ell+1}_k].
		\end{equation}
		The sum of correlations at level $\ell$ is given by,
		\begin{equation}
		I[\rho,\mathbf{X}^{\ell}] \stackrel{\mathrm{def}}{=} \sum_{k=1}^{n_{\ell}}I[\rho,\mathbf{X}^{\ell}_k] = \sum_{k=1}^{n_{\ell}}\left(\sum_{i=1}^{n^k_{\ell}}S[\rho,\mathbf{X}_{k_i}^{\ell}] - S[\rho,\mathbf{X}^{\ell}_k]\right),
		\end{equation}
		so that using (\ref{identify}) the sum of the correlations at level $\ell$ and level $(\ell+1)$ becomes,
		\begin{align}
		I[\rho,\mathbf{X}^{\ell+1}] + I[\rho,\mathbf{X}^{\ell}] &= \sum_{k=1}^{n_{\ell+1}}I[\rho,\mathbf{X}^{\ell+1}_k] + \sum_{k=1}^{n_{\ell}}I[\rho,\mathbf{X}^{\ell}_k]\nonumber\\
		&= \sum_{k=1}^{n_{\ell+1}}\left(\sum_{i=1}^{n^k_{\ell+1}}S[\rho,\mathbf{X}^{\ell+1}_{k_i}]\right) - \sum_{k=1}^{n_{\ell}}S[\rho,\mathbf{X}_{k}^{\ell}],\label{IplusI2}
		\end{align}
		where we used from (\ref{identify}) and (\ref{intersect}) that,
		\begin{equation}
		\sum_{k=1}^{n_{\ell}}\left(\sum_{i=1}^{n_{\ell}^k}S[\rho,\mathbf{X}^{\ell}_{k_i}]\right) = \sum_{k=1}^{n_{\ell+1}}S[\rho,\mathbf{X}_{k}^{\ell+1}].
		\end{equation}
		Summing over all $\ell$ up to some $\ell_n$ beginning with $\mathbf{X}^0 \stackrel{\mathrm{def}}{=} \mathbf{X}$, the expression in (\ref{IplusI2}) generalizes to,
		\begin{equation}
		\sum_{\ell=0}^{\ell_n}I[\rho,\mathbf{X}^{\ell}] = \sum_{k=1}^{n_{\ell_n}}\left(\sum_{i=1}^{n^k_{\ell_n}}S[\rho,\mathbf{X}^{\ell_n}_{k_i}]\right) - S[\rho,\mathbf{X}].
		\end{equation} 
		If one continues splitting until each subset $\mathbf{X}^{\ell_N}_{k_i} = \mathbf{X}_i$ contains only a single variable from $\mathbf{X}$, then the above eq. becomes,
		\begin{equation}
		\sum_{\ell=0}^{\ell_N}I[\rho,\mathbf{X}^{\ell}] = \sum_{i=1}^NS[\rho,\mathbf{X}_i] - S[\rho,\mathbf{X}] = I[\rho],
		\end{equation}
		which is the total correlation.  Given that the subsets at each split level are disjoint (\ref{identify}), then the above equation is independent of the choice of splitting.  This proves the \hyperref[watanabe]{theorem}.
	\end{proof}
	
	\section{Consequences of the derivation}\label{consequences}
	Here we will analyze some basic properties of total correlation and mutual information, as well as show their consistency with the design criteria.  
	
	\subsection{Inferential transformations (again)}
	Due to the results of our design derivation, we can quantify how the amount of correlations change with the inferential transformations from \hyperref[class]{Section II} and obtain a better understanding of them.  We will also discuss some of the useful properties of the mutual information in subsections \hyperref[rednoise]{6.2} and \hyperref[data]{6.3}.
	\subsubsection{Type \textbf{I}: Coordinate transformations (again)}
	Consider \hyperref[coordtrans]{type \textbf{I}} transformations, which are coordinate transformations.  Under type \textbf{I}, the density changes from $p(x_1,\dots,x_N)$ to $p'(x'_1,\dots,x'_N)$ so that the probabilities remain equal,
	\begin{equation}
	p(x_1,\dots,x_N)dx = p'(x'_1,\dots,x'_N)dx'.\label{split}
	\end{equation}
	The individual density $p(x_1,\dots,x_N)$ transforms as,
	\begin{equation}
	p(x_1,\dots,x_N) = p'(x'_1,\dots,x'_N)\gamma(x'_1,\dots,x'_N).
	\end{equation}
	If the transformation (\ref{split}) is split coordinate invariant, then we have that the new marginals must obey,
	\begin{equation}
	p(x_i)dx_i = p'(x'_i)dx'_i.
	\end{equation}
	It is important to note that the joint density $p'(x')$ and the marginals $p'(x'_i)$ are necessarily different functions than $p(x)$ and $p(x_i)$ and are not simply the same function defined over the transformed space, which would be written as $p(x')$ and $p(x'_i)$.  Due to (\ref{transforms}), it is true that, 
	\begin{align}
	I[\rho',\mathbf{X}'] &= \int dx'\, p'(x'_1,\dots,x'_N)\log\frac{p'(x'_1,\dots,x'_N)}{\prod_{i=1}^Np'(x'_i)}\nonumber\\
	&= \int dx\, p(x_1,\dots,x_N)\log\frac{p(x_1,\dots,x_N)}{\prod_{i=1}^Np(x_i)} = I[\rho,\mathbf{X}],
	\end{align}
	which is coordinate invariant since the Jacobian factors $\gamma(x')$ cancel in the logarithm.
	\subsubsection{Type \textbf{II}: Entropic updating (again)}
	We can determine how the amount of correlation changes under \hyperref[entropicupdating]{type \textbf{II}} transformations.  One can use the relative entropy (\ref{entropy}) to update a joint prior distribution $q(x_1,\dots,x_N)$ to a posterior distribution $p(x_1,\dots,x_N)$ when information comes in the form of constraints,
	\begin{equation}
	\langle f(x_1,\dots,x_N) \rangle = \int dx\, p(x_1,\dots,x_N)f(x_1,\dots,x_N) = \kappa,\label{constraint}
	\end{equation}
	where $f(x_1,\dots,x_N)$ is a generic function of the variables and $\kappa$ is an arbitrary constant.  Maximizing (\ref{entropy}) with respect to the constraint (\ref{constraint}) leads to,
	\begin{equation}
	p(x_1,\dots,x_N) = \frac{q(x_1,\dots,x_N)}{Z}\exp\left[-\beta f(x_1,\dots,x_N)\right],
	\end{equation}
	where $Z = \int dx\, q(x_1,\dots,x_N)\exp\left[-\beta f(x_1,\dots,x_N)\right]$ and $\beta$ is a Lagrange multiplier.  The marginals of the updated distribution $p(x_1,\dots,x_N)$ are,
	\begin{equation}
	p(x_i) = \frac{q(x_i)}{Z}\int d\bar{x}_i\, q(x_i|\bar{x}_i)e^{-\beta f(x_1,\dots,x_N)},
	\end{equation}
	where $ d\bar{x}_i = \prod_{k \neq i}dx_k$ and $\bar{x}_i = \mathbf{X}\backslash\mathbf{X}_i$ (i.e. the total space $\mathbf{X}$ without the variable $\mathbf{X}_i$).  The total correlation of the prior is,
	\begin{equation}
	I[\rho,\mathbf{X}] = \int dx\, q(x_1,\dots,x_N)\log\frac{q(x_1,\dots,x_N)}{\prod_{i=1}^Nq(x_i)},
	\end{equation}
	while the total correlation of the posterior is,
	\begin{align}
	I[\rho',\mathbf{X}] &= \int dx\, p(x_1,\dots,x_N)\log\frac{p(x_1,\dots,x_N)}{\prod_{i=1}^Np(x_i)},\nonumber\\
	&= \int dx\, \frac{q(x_1,\dots,x_N)}{Z}\exp\left[-\beta f(x_1,\dots,x_N)\right]\log\frac{q(x_1,\dots,x_N)}{\prod_{i=1}^Nq(x_i)}\nonumber\\
	& -\left\langle \beta f(x_1,\dots,x_N) + \log\Big(Z\frac{\prod_{i=1}^Np(x_i)}{\prod_{i=1}^Nq(x_i)}\Big)\right\rangle_{p(x_1,\dots,x_N)} .
	\end{align}
	Transformations of \hyperref[entropicupdating]{type \textbf{II}} retain some of the correlations from the prior into the posterior. The amount of correlations may increase or decrease depending on $f(x_1,\dots,x_N)$. Note that even if $f(x_1,\dots,x_N)=f(x_i)$, that the amount of correlations can still change because, although $q(x_k|\bar{x}_k)$ remains fixed, $q(x_i)\rightarrow p(x_i)$ becomes redistributed in a way that may group on highly correlated areas or not. 
	\subsubsection{Type \textbf{III}: Marginalization (again)}\label{margagain}
	For \hyperref[marginalization]{type \textbf{III}} transformations, we can determine the difference in the amount of correlations when we marginalize over a set of variables.  Consider the simple case where $\mathbf{X} = \mathbf{X}_1\times\mathbf{X}_2\times\mathbf{X}_3$ consists of three variables.  The amount of correlations between these two sets can be written using the grouping property,
	\begin{align}
	I[\rho,\mathbf{X}] &= \int dx_1dx_2\,p(x_1,x_2)\log\frac{p(x_1,x_2)}{p(x_1)p(x_2)}\nonumber\\
	& + \int dx_1dx_2dx_3\,p(x_1,x_2)p(x_3|x_1,x_2)\log\frac{p(x_3|x_1,x_2)}{p(x_3)}\nonumber\\
	I[\rho,\mathbf{X}_1;\mathbf{X}_2;\mathbf{X}_3] &= I[\rho,\mathbf{X}_1;\mathbf{X}_2] + I[\rho,(\mathbf{X}_1\times\mathbf{X}_2);\mathbf{X}_3]\label{cond},
	\end{align}
	where each term on the right hand side is a mutual information.  The first term on the right hand side is the mutual information between the variables $\mathbf{X}_1$ and $\mathbf{X}_2$ while the second is the mutual information between the joint space $(\mathbf{X}_1\times\mathbf{X}_2)$ and the space $\mathbf{X}_3$, which quantifies the correlations between the third variable and the rest of the proposition space.  Essentially, (\ref{cond}) shows that we can count correlations in pairs of subspaces provided we don't \textit{overcount}, which is the reason why the joint space $(\mathbf{X}_1\times\mathbf{X}_2)$ appears in the second term.\\
	\indent  Thus, if one marginalizes over $\mathbf{X}_3$ so that the joint space becomes $\mathbf{X}_1\times\mathbf{X}_2$, then the global correlations lost are given by the difference from (\ref{cond}),
	\begin{equation}
	\Delta I = I[\rho,\mathbf{X}_1;\mathbf{X}_2] = I[\rho,\mathbf{X}_1;\mathbf{X}_2;\mathbf{X}_3] - I[\rho,(\mathbf{X}_1\times\mathbf{X}_2);\mathbf{X}_3].
	\end{equation}
	\indent  We can break down mutual informations further by using the grouping property again, which leads to the second term in the above relation becoming,
	\begin{align}
	I[\rho,(\mathbf{X}_1\times\mathbf{X}_2);\mathbf{X}_3] &= \int dx_1dx_3\, p(x_1,x_3)\log\frac{p(x_1,x_3)}{p(x_1)p(x_3)}\nonumber\\
	&+ \int dx_1dx_2dx_3\, p(x_1)p(x_2,x_3|x_1)\log\frac{p(x_2,x_3|x_1)}{p(x_2|x_1)p(x_3|x_1)}\nonumber\\
	&= I[\rho,\mathbf{X}_1;\mathbf{X}_3] + I[\rho,\mathbf{X}_2;\mathbf{X}_3|\mathbf{X}_1],
	\end{align}
	where the quantity,
	\begin{equation}
	I[\rho,\mathbf{X}_2;\mathbf{X}_3|\mathbf{X}_1] = \int dx_1\, p(x_1)\int dx_2dx_3\,p(x_2,x_3|x_1)\log \frac{p(x_2,x_3|x_1)}{p(x_2|x_1)p(x_3|x_1)},
	\end{equation}
	is typically called the \textit{conditional} mutual information (CMI) \cite{CoverThomas_Book}.  In general we have the chain rule,
	\begin{equation}
	I[\rho,\mathbf{X}_1,\dots,\mathbf{X}_n;\mathbf{Y}] = \sum_{i=1}^nI[\rho,\mathbf{X}_i;\mathbf{Y}|\mathbf{X}_{i-1},\dots,\mathbf{X}_{1}].\label{chain}
	\end{equation}Thus if one begins with the full space $p(x,y) = p(x_1,x_2,y)$ and marginalizes over $\mathbf{X}_2$,
	\begin{equation}
	p(x_1,y) = \int dx_2\, p(x_1,x_2,y)\label{marg},
	\end{equation}
	then the global correlations lost are given by the CMI,
	\begin{equation}
	\Delta I = I[\rho,\mathbf{X};\mathbf{Y}] - I[\rho,\mathbf{X}_1;\mathbf{Y}] = I[\rho,\mathbf{X}_2;\mathbf{Y}|\mathbf{X}_1].
	\end{equation}
	Such a marginalization leaves the correlations invariant whenever the CMI is zero, i.e. whenever $p(x_2,y) = p(x_2)p(y)$ are independent.

	\subsubsection{Type \textbf{IV}: Products (again)}\label{prodagain}
	We find that transformations of \hyperref[products]{type \textbf{IV}} give the same expression.  Consider the reverse of (\ref{marg}) in which,
	\begin{equation}
	p(x_1,y) = p(x_1)p(y|x_1) \xrightarrow{\mathbf{IV}} p(x_1,x_2,y) = p(x_1,x_2)p(y|x_1,x_2).
	\end{equation}
	Then the change in mutual information is given by the same expression as (\ref{cond}), so that the \textit{gain} in global correlations is simply the value of the CMI.  
	
	\subsection{Type \textbf{IVa} and \textbf{IVb}: Redundancy and noise}\label{rednoise}
	Using the special case of mutual information, we can better analyze and interpret the two special cases of \hyperref[products]{type \textbf{IV}} that we introduced in \hyperref[class]{Section II}. We can show how the amount of bipartite correlations change with the addition or subtraction of variables exhibiting \textit{redundancy} or \textit{noise}, which are the special cases \textbf{IVa} and \textbf{IVb}, respectively.
	
	Consider the two subspaces $\mathbf{X}$ and $\mathbf{Y}$.  If the subspace $\mathbf{X} = \mathbf{X}_1\times\dots\times\mathbf{X}_N$ is a collection of $N$ variables, then the joint distribution can be written
	\begin{equation}
	p(x,y) = p(x_1,\dots,x_N,y) = p(x_1,\dots,x_N)p(y|x_1,\dots,x_N).
	\end{equation}
	If the conditional probability $p(y|x_1,\dots,x_N)$ is independent of $\mathbf{X}_i$ but the distribution $p(x_i|x_1,\dots,x_{i-1},x_{i+1},\dots,x_N)$ is not, then we say that $\mathbf{X}_i$ is \textit{redundant}. This is equivalent to the condition in (\ref{iva}) whenever $x_i = f(\mathbf{X}\backslash\mathbf{X}_i)$. In this case, the amount of correlations on the full set of variables $I[\rho]$ and the set without $\mathbf{X}_i$, ($\tilde{\mathbf{X}} = \mathbf{X}\backslash\mathbf{X}_i$), are equivalent, 
	\begin{align}
	p(x,y) &= p(x_1,\dots,x_{i-1},x_{i+1},\dots,x_N)\nonumber\\
	&\times p(y|x_1,\dots,x_{i-1},x_{i+1},\dots,x_N)\nonumber\\
	&\times\delta(x_i - f(x_1,\dots,x_{i-1},x_{i+1},\dots,x_N))\nonumber\\
	&\Rightarrow I[\rho,\mathbf{X}] = I[\tilde{\rho},\tilde{\mathbf{X}}].\label{redundant2}
	\end{align}
	Hence, the bipartite correlations in $(\mathbf{X}_i,\mathbf{Y})$ are redundant.  While the condition that $\mathbf{X}\backslash\mathbf{X}_i$ leads to the same mutual information as $\mathbf{X}$ can be satisfied by  (\ref{redundant2}), it is not necessary that $\mathbf{X}_i = f(\tilde{\mathbf{X}})$.  In an extreme case, we could have that $\mathbf{X}_i$ is independent of both $\mathbf{X}$ and $\mathbf{Y}$,
	\begin{align}
	p(x,y) &= p(x_i)p(x_1,\dots,x_{i-1},x_{i+1},\dots,x_N)\nonumber\\
	&\times p(y|x_1,\dots,x_{i-1},x_{i+1},\dots,x_N)\label{noise}.
	\end{align}
	In this case we say that the variable $\mathbf{X}_i$ is \textit{noise}, meaning that it adds dimensionality to the space $(\mathbf{X}\times\mathbf{Y})$ without adding correlations.  In the redundant case, the variable $\mathbf{X}_i$ does \textit{not} add dimension to the manifold $(\mathbf{X}\times\mathbf{Y})$. 
	
	In practice, each set of variables $(\mathbf{X},\mathbf{Y})$ will contain some \textit{amount} of redundancy and some \textit{amount} of noise.  We could always perform a coordinate transformation that takes $\mathbf{X} \rightarrow \mathbf{X}'$ and $\mathbf{Y}\rightarrow \mathbf{Y}'$ where,
	\begin{align}
	\mathbf{X}' &= \mathbf{X}_{\mathrm{red}}\times\mathbf{X}_{\mathrm{noise}}\times\mathbf{X}_{\mathrm{corr}}\nonumber\\
	\mathbf{Y}' &= \mathbf{Y}_{\mathrm{red}}\times\mathbf{Y}_{\mathrm{noise}}\times\mathbf{Y}_{\mathrm{corr}},
	\end{align}
	where $\mathbf{X}_{\mathrm{red}},\mathbf{X}_{\mathrm{noise}}$ are redundant and noisy variables respectively and $\mathbf{X}_{\mathrm{corr}}$ are the parts left over that contain the relevant correlations.  Then the joint distribution becomes,
	\begin{align}
	p(x',y') &= p(x_{\mathrm{noise}})p(y_{\mathrm{noise}})p(x_{\mathrm{corr}})p(y_{\mathrm{corr}}|x_{\mathrm{corr}})\nonumber\\
	&\times\delta(x_{\mathrm{red}} - f(x_{\mathrm{corr}}))\delta(y_{\mathrm{red}} - g(y_{\mathrm{corr}})).
	\end{align}
	Thus we have that,
	\begin{align}
	I[\rho,\mathbf{X};\mathbf{Y}] &= I[\rho',\mathbf{X}_{\mathrm{red}}\times\mathbf{X}_{\mathrm{noise}}\times\mathbf{X}_{\mathrm{corr}};\mathbf{Y}_{\mathrm{red}}\times\mathbf{Y}_{\mathrm{noise}}\times\mathbf{Y}_{\mathrm{corr}}]\nonumber\\
	&= I[\rho',\mathbf{X}_{\mathrm{corr}};\mathbf{Y}_{\mathrm{corr}}].
	\end{align}
	These types of transformations can be exploited by algorithms to reduce the dimension of the space $(\mathbf{X}\times\mathbf{Y})$ to simplify inferences.  This is precisely what machine learning algorithms are designed to do \cite{Carrara_Ernst}.  One particular effort to use mutual information directly in this way is the \textit{Information Bottleneck Method} \cite{Bottleneck}.
	
	\subsection{The data processing inequality}\label{data}
	The data processing inequality is often demonstrated as a consequence of the definition of MI.  Consider the following Markov chain,
	\begin{equation}
	\Theta \rightarrow \mathbf{X} \rightarrow \mathbf{Y} \quad \Rightarrow \quad p(\theta,x,y) = p(\theta)p(x|\theta)p(y|x) = p(\theta)p(x|\theta)\delta(y - f(x))\label{markov}.
	\end{equation}
	We can always consider the MI between the joint space $(\mathbf{X}\times\mathbf{Y})$ and $\Theta$, $I[\rho,\mathbf{X}\times\mathbf{Y};\Theta]$, which decomposes according to the chain rule (\ref{chain}) as,
	\begin{align}
	I[\rho,\mathbf{X}\times\mathbf{Y};\Theta] &= I[\rho',\mathbf{X},\Theta] + I[\rho'',\mathbf{Y};\Theta|\mathbf{X}]\nonumber\\
	&= I[\rho''',\mathbf{Y};\Theta] + I[\rho'''',\mathbf{X};\Theta|\mathbf{Y}].
	\end{align}
	The conditional MI $I[\rho'',\mathbf{Y};\Theta|\mathbf{X}]$ is however,
	\begin{equation}
	I[\rho'',\mathbf{Y};\Theta|\mathbf{X}] = \int dx\, p(x) \int dyd\theta\, p(y,\theta)\log\frac{p(\theta|y,x)p(y|x)}{p(\theta|x)p(y|x)},
	\end{equation}
	which is zero since $p(\theta|y,x) = p(\theta|x)$.  Since MI is positive, we then have that for the Markov chain (\ref{markov}),
	\begin{equation}
	I[\rho,\mathbf{X};\Theta] \geq I[\rho',\mathbf{Y};\Theta],\label{dpi}
	\end{equation}  
	of which now we can interpret as expressing a loss of correlations.
	Equality is achieved only when $I[\rho,\mathbf{X};\Theta|\mathbf{Y}]$ is also zero, i.e. when $\mathbf{Y}$ is a sufficient statistic for $\mathbf{X}$.  We will discuss the idea of sufficient statistics in \hyperref[sufficiency]{Section VII}. 
	
	\subsection{Upper and lower bounds for mutual information}
	The upper bounds for mutual information can be found by using the case of complete correlation, $y = f(x)$,
	\begin{align}
	I[\rho,\mathbf{X};\mathbf{Y}] 
	&= \int dxdy\,p(x)\delta(y - f(x))\log\frac{p(x|y)}{p(x)}\nonumber\\
	&= -\int dx\, p(x)\log\frac{p(x)}{p(x|f(x))} = S[p(x),p(x|f(x))]\label{upper_1},
	\end{align}
	which is the relative entropy of $p(x)$ with respect to $p(x|f(x))$.  If $f(x)$ is a coordinate transformation, i.e. is a bijection, then eq. (\ref{upper_1}) becomes,
	\begin{equation}
	I[\rho,\mathbf{X};\mathbf{Y}] = \int dx\, p(x)\log\frac{\delta(x - f^{-1}f(x))}{p(x)} = \infty,
	\end{equation}
	since $\delta(x - f^{-1}f(x)) = \delta(0) = \infty$.  Hence, the MI is unbounded from above in the continuous case.  In the discrete case \cite{Merkh} we find,
	\begin{align}
	I[\rho,\mathcal{X};\mathcal{Y}] &= \sum_{x_i,y_j}P(x_i)\delta_{yx}\log\frac{P(x_i|y_j)}{P(x_i)}\nonumber\\
	&= -\sum_{x_i} P(x_i)\log P(x_i) = H[\rho,\mathcal{X}]\label{discrete_2},
	\end{align}  
	where $H[\rho,\mathcal{X}]$ is the Shannon entropy and the notation $\mathcal{X} \subset \mathbf{X}$ refers to a \textit{sample} drawn from the ambient space $\mathbf{X}$.  Since (\ref{discrete_2}) does not depend on the functional form of $f$, the upper bound is simply the Shannon entropy of one of the two variables.  This can be seen by expanding the discrete MI as a sum of Shannon entropies,
	\begin{equation}
	I[\rho,\mathcal{X};\mathcal{Y}] = H[\rho,\mathcal{X}]  +H[\rho,\mathcal{Y}] - H[\rho,\mathcal{X},\mathcal{Y}],
	\end{equation}
	which in the case of complete correlation ($y = f(x)$), the joint Shannon entropy becomes,
	\begin{align}
	H[\rho,\mathcal{X};\mathcal{Y}] &= -\sum_{x_i,y_j}P(x_i)\delta_{yx}\log(P(x_i)\delta_{yx})\nonumber\\
	&= -\sum_{x_i}P(x_i)\log P(x_i) = H[\rho,\mathcal{X}],
	\end{align}
	and so the upper bound is,
	\begin{equation}
	I_{\max}[\rho,\mathcal{X};\mathcal{Y}] = \max \left\{H[\rho,\mathcal{X}],H[\rho,\mathcal{Y}]\right\}.
	\end{equation}
	If $y = f(x)$ is a bijection, then the entropies $H[\rho,\mathcal{X}] = H[\rho,\mathcal{Y}]$ since $\mathcal{Y}$ is just a reparameterization of $\mathcal{X}$ and hence the probabilities $P(x) = P(y)$.\\
	\indent  Since the mutual information is unbounded from above, then so too is the total correlation due to the additivity in the chain rule (\ref{cond}).
\end{appendix}

\bibliography{bigbib}{}
\bibliographystyle{plain}

\end{document}